\documentclass[final,5p,times,twocolumn]{elsarticle}

\newcounter{prob_num}
\newcounter{MYtempeqncnt}
\usepackage{amssymb}
\usepackage{amsmath}
\usepackage{graphicx}

\begin{document}

\begin{frontmatter}

\title{DSAT-MAC : Dynamic Slot Allocation based TDMA MAC protocol for Cognitive
Radio Networks}

\author{Rajeev K. Shakya, Satyam Agarwal, Y. N. Singh, Nishchal K. Verma, and
Amitabha Roy}

\address{Department of Electrical Engineering\\
Indian Institute of Technology, Kanpur, India\\
Email: \{rajeevs, satyamag, ynsingh,
nishchal, aroy\}@iitk.ac.in}

\begin{abstract}
Cognitive Radio Networks (CRN) have enabled us to efficiently reuse the
underutilized
radio spectrum. 
The MAC protocol in CRN defines the spectrum usage by sharing the channels
efficiently
among users. 
In this paper we propose a novel TDMA based MAC protocol
with dynamically allocated slots. Most of the MAC
protocols proposed in the
literature employ Common Control Channel (CCC) to manage the resources among
Cognitive Radio (CR) users. 
Control channel saturation in case of large number of CR users is one of the
main drawbacks of the CCC based MAC protocols.
In contrast with CCC based MAC protcols, DSAT-MAC protocol is based on
the TDMA mechanism,
without using any CCC for control information exchange. The channels are divided
into time slots and CR users send their control or data packets over their
designated slot. The protocol ensures that no slot is left vacant. This
guarantees full use of the available spectrum. The protocol includes the
provision for Quality of Service, where real-time and safety critical data is
transmitted with 
highest priority and least delay. The protocol also ensures a fair sharing of
available spectrum among the CR users, with the mechanism to regulate the
transmission of malicious nodes. Energy saving techniques are also presented for longer life of battery operated CR nodes. 
Theoretical analysis and simulations over ns-2 of
the proposed protocol reveal that the protocol performs better in various CR
adhoc network applications.

\end{abstract}

\end{frontmatter}
%%%%%%%%%%%%%%%%%%%%%%%%%%%%%%%%%%%%%%%%%%%%%%%%%%
%%%%%%%%%%%%%%%%%%%%%%%%%%%%%%%%%%%%%%%%%%%%%%%%%%
\section{Introduction}
Increasing usage and diverse applications of the wireless communication networks
have led to a high demand of vacant spectrum. Most of the wireless communication
is carried out on the licensed band. Many studies on spectrum utility like that
in [1] have revealed that these licensed spectrum are being significantly
underutilized. Joseph Mitola in 1999 [2] devised a unique method to use the
unused spectrum by introducing the concept of Cognitive Radio. Cognitive Radio
controls its transmitter and receiver parameters intelligently so as to use the
vacant licensed spectrum without affecting the working of the primary licensed
users (PU) of these licensed bands. Firstly the Cognitive Radio users (CR
users) sense for vacant spectrum in their surroundings and then carry on
communications among other CR users. The CR users are permitted to use these
spectrum provided they do not cause any harmful interference to the PU and allow
fair share of the available spectrum among themselves. Television Broadcast
frequencies below 700 
MHz have been proposed for the CR operations [3].\\
Spectrum sensing and channel assignment are the keys behind the successful
operation of Cognitive Radio Networks (CRNs) formed by these CR nodes. 
Spectrum sensing is employed at the physical layer and is responsible for
detecting any PU activity and locating white spaces (vacant bands) in the
spectrum.
Channel
allocation and management is mainly carried out by the MAC protocol at the link
layer. Most of the MAC protocols proposed in the literature utilize a Common
Control Channel (CCC) to exchange control information among the CR nodes 
and assign the vacant channels to various nodes according to channel
negotiations in the network. 
Generally, a control channel is
one of the unlicensed band assigned solely for exchanging control information.
The protocols proposed in [4][5] have employed CCC for exchanging control
information and negotiating the vacant bands for their data transmissions.

\subsection{Cognitive Radio}
Cognitive Radio (CR) is a key enabling technology for the Dynamic Spectrum Access (DSA). The radio senses its surrounding environment for the vacant spectrum which is not being utilised and makes use of it to transmit information to other cognitive radio nodes. Hence it controls its transmitter or receiver parameters to intelligently adapt to its surroundings without affecting the working of the primary licensed users. The vacant spectrum are also referred to as Spectrum Holes or White Spaces. The cognition power of the radio i.e. the power of the radio to adapt itself with its surrounding makes it special. This capability of the cogntive radio to use and share the spectrum in an opportunistic way is called Bandwidth Harvesting.\\
\indent There are many hardware and software developments being carried out to
facilitate the growth of cognitive radio. The concept of Software Defined Radio
(SDR) given by Joseph Mitola has now became an exciting new field of study. By
the SDR, the radio could be programmed and used as desired. The built-in
software component in the SDR imparts cognitive ability to the radio and makes
it smart. The transmitter and receiver frequencies can suitably be varied using
this software radio. This enables us to properly use  the white spaces, which
results in increasing spectrum utility. There are many applications that could
be introduced using this concept without any problem of spectrum availability.
Some of the main components of the cognitive radio are as follows: 
\begin{itemize}
\item \textbf{Primary Users} : Also called as Licensed Users. These users have a license to operate in a certain spectrum band.  
\item \textbf{Primary Network} : This is the network on which the primary users operate. This is an existing network infrastructure and has all rights over a particular spectrum band.
\item \textbf{Secondary Users} : Also called Unlicensed Users or Cognitive Users. They do not hold a licence for operating in the particular band. They can only use the spectrum when it is free or not used by the primary users.
\item \textbf{Secondary Network} : Also called Cognitive Radio Network. The network formed by the secondary nodes with no license to operate in the desired spectrum band.
\end{itemize}
\indent The Cognitive Radio Network (CRN) is formed by a group of nodes with
cognitive radio capabilities and which operate on opportunistically available
licensed spectrum. The CRN should work in such a way that they do not cause any
harmful interference to the primary users. Hence it is very much required in the
CRNs that the nodes regulate their transmission power, the spectrum band on
which they transmit/receive and the time as to when to transmit. The cognitive
users have to sense the spectrum to locate the spectrum holes or white spaces
and then they negotiate with the other CR users for accessing the channel. After
every certain amount of time they have to sense the channel for any primary user
activity so that they do not hamper the communication of the primary users.
Television Broadcast frequencies below 700MHz have been proposed for CR
operations[6]. In addition to this the 
spectrum reserved for public service operations can also be used with frequent
Quiet periods for CR operations. 
\subsection{Classification of Cognitive Radio Networks}
The Cognitive Radio Networks could be classified into two main categories namely:
\begin{itemize}
\item \textbf{Centralised Cognitive Radio Networks}: In this type of network there is one main base station and multiple CR users which coordinate with the base station. The base station determines the spectrum band which are empty and allocate the bands to the different CR users according to their requirements. The CR users send their sensing parameters to the base station to determine which all spectrum bands to be used. The base station may also be connected to a central spectrum server which could provide the information regarding the availability of the spectrum bands.
\item \textbf{Distributed Cognitive Radio Networks} : In this type of network there is no central controlling unit, while all the CR users are independent. A good coordination i.e. MAC protocol is required to coordinate among them. Each CR user has a sensing module which senses the channel for white spaces and negotiates the band with the other users in the network. These are also called Cognitive Radio Adhoc Network.
\end{itemize}

\indent Some aspects of Cognitive Radio Networks are as follows :
\begin{itemize}
\item \textbf{Spectrum Sensing} : Spectrum sensing implies detecting the unused spectrum and utilising the spectrum without causing harmful interference to primary users.
\item \textbf{Spectrum Management} : This means to manage the available spectrum resources and using the best spectrum to meet the communication requirements.
\item \textbf{Spectrum Mobility} : This applies to maintaining unhindered communication even while transiting from one spectrum band to a better spectrum band.
\item \textbf{Spectrum Sharing} : Sharing the available spectrum fairly among other CR users.
\end{itemize}
\indent With the advancement in the Cognitive Radio Networks the IEEE 802.22 standards for the Wireless Regional Area Networks has been formulated. This is for the infrastructure based Cognitive Radio Networks with the central base station managing the spectrum and allocating resources to other users.

\subsection{Application of Cognitive Radio Networks}
There are many emerging cognitive radio network applications given in [4]. Following illustrate some of these applications.
\begin{itemize}
\item \textbf{Public Safety System} : Wireless communications are extensively used for emergency services like police and medical services to respond to emergency situations. These safety workers are equipped with various modules like mobiles, video telephony etc. to improve their efficiency and to be always be in touch with their co-workers and central authorities. The CRNs are extensively proposed for these kinds of services as they promise high spectrum coverage in such situations where large bandwidth might be required.
\item \textbf{Smart Grid Networks} : A smart grid is a network which attempts to intelligently guess the behaviour of all electric power users and smartly respond to their actions in order to efficiently deliver reliable, economic, and sustainable electricity service. This increases the reliability and efficiency of electricity transmission and reduces cost for consumers and producers. These grids make use of WiFi, ZigBee, TV White Spaces (TVWS), cellular networks in rural areas as the primary spectrum for their cognitive nodes for easy communication.
\item \textbf{Wireless Medical Network} : There is a large variety of applications and services that are used in the hospitals like online monitoring of patients health, online medical records of patients. These CRNs could prove very useful at these places where the unlicensed spectrums are used and spectrum demands are high.
\item \textbf{Cognitive Radio Sensor Networks} : Sensor networks practically generate bursty data. The transmission of such bursty data at any given time requires a high bandwidth for a small period of time. Else small bandwidth or less number of channels may cause high collisions and make the system inefficient[5]. Realising CRSNs involve few challenges such as design of power-efficient and low-cost cognitive radio sensor nodes, and opportunistic multi-hop routing over licensed and unlicensed spectrum bands. These could be used for real-time applications, multimedia applications or indoor sending applications. 
\item \textbf{Other Areas}: Cognitive Radio Networks would prove efficient in areas where the main spectrum used is the ISM unlicensed bands. As there is large number of applications coming up in this part of the band, hence there is a scarcity of unlicensed spectrum to communicate. CRNs are highly capable in covering these shortcomings in various such applications.
\end{itemize}

\subsection{Motivation}
The success of Cognitive Radio networks highly depends on how the spectrum is being accessed. There has been great interest in MAC protocols in recent times. The Medium Access Control (MAC) forms an essential component behind the successful and efficient working of the CRNs. The MAC provides the rules for assessing the channel and how to share the available channels among other CR users.\\
\indent Mostly the MAC for CRNs are based on using a dedicated control channel [10][11] for suitable exchange of control informations. This kind of mechanism have limitations like channel saturation and availability of such channel. There are other methods proposed as well in the literature like hopping based control channels [13] and CSMA based MAC protocols but they too have their drawbacks like unnecessary delays.\\
\indent  Hence while the MAC protocol as a high contribution inefficient working of CRNs, the current protocols proposed have many limitations. Motivated by this problem, we have tried to develop a new MAC protocol which could achieve optimal performance. 

\subsection{Main Contributions}
\indent In this article we propose a novel TDMA based MAC protocol for cognitive
radio adhoc networks. We have introduced a scheme to guarantee QoS, wherein the
real-time traffic is given high priority as compared to non-real time traffic.
We have also proposed a technique to restrict the operation of malicious user
trying to hog up the bandwidth. The protocol aims to achieve the following.
\begin{itemize}
\item Designing an effective method to exchange control information among CR nodes without the use of a dedicated control channel.
\item Maintain a level of QoS among the CR users and giving priority to safety critical data and real-time traffic.
\item Maintaining fairness among all the CR users and giving equal chance to all the nodes to access the available channel.
\item The protocol aims at increasing the utility of the spectrum by fully utilising the available spectrum.
\item Making the system robust against any operation of malicious user.
\item The protocol should guide the system efficiently and coordinate all the users effectively without the help of any master node.
\item The CR nodes could be battery operated so the protocol should be energy-efficient.
\end{itemize}
\indent We have analysed our protocol by simulating in ns-2. We have shown the effect on performance of the protocol under varying conditions. We have also give a theoretical derivation of the maximum throughput of the system and energy efficiency achieved by the protocol.
% \subsection{Thesis Organisation}

The rest of the article is organised as follows. Section 2 of this article gives
a brief description of the existing MAC protocols for the Cognitive Radio
Networks. Section 3 describes the detailed outline of the proposed MAC protocol.
Section 4 gives a detailed analysis of the protocol with the analytical and
simulation results. Section 5 concludes the article with the list of future
works for research on CR Medium Access Control protocols.

%%%%%%%%%%%%%%%%%%%%%%%%%%%%%%%%%%%%%%%%%%%%%%%%%%%
%%%%%%%%%%%%\input{paperch2}%%%%%%%%%%%%%%%%%%%%%%%%%%%%%
%%%%%%%%%%%%%%%%%%%%%%%%%%%%%%%%%%%%%%%%%%%%%%%%%%%
% \section{Literature Survey}
\section{Related Work}
The MAC protocol directs the CR users that which user would access the spectrum
and when to access the channel. In this section we would discuss the existing
MAC protocols for the Cognitive Radio Networks. The MAC protocol is
designed keeping in mind the application and limitations of the network. There
are different MAC protocols for Centralised CRN and Distributed or Adhoc CRN. In
the former case it utilises the base station to coordinate the various
activities and resources among CR users while in the latter all the CR users
work independently and share the resources among them intelligently. Assume a
scenario in which the CR users have high mobility the time synchronisation
becomes difficult. The MAC protocol in such cases have to be adjusted
accordingly. Thus the operation of MAC protocol must be adaptive to suit
specific application or specific environment, hence it makes standardization of
protocol challenging.\\
\indent The MAC protocol for CRNs is different from the classical MAC protocol in the way that the MAC in CRN is closely coupled with the physical layer of the device. This is because for the MAC to efficiently manage the channel, needs all the related information regarding the channel status. Based on the RF status from the physical layer, the MAC decides the sensing time and the schedule of transmission. MAC layer is also responsible for the duration of the optimal sensing and transmission times and the method of searching for the vacant spectrum such that the performance is optimal[8]. Channel has to be sensed at regular intervals to detect for any primary user activity. It is clear that high sensing time ensures correct detection of the PU activity while reduces throughput whereas the large transmission time guarantees high throughput while increasing the probability of false alarm and result in unwanted interference with the primary users. Hence an optimal solution is suggested in [9] based on numerical 
optimization.\\
\indent The MAC protocols are generally based on the type of cognitive radio networks i.e. the MAC protocols can primarily be divided into MAC for Centralised CRN and MAC for Adhoc CRN. The MAC protocols could also be divided into time slotted or random access or hybrid protocol depending upon the type of its working. One of the main challenges in the spectrum access is control channel assignment. Control channel is used for the CR users to coordinate among themselves and exchange control messages. This channel must be free from any interruption by the PUs. One solution is to allocate an unlicensed channel for transmitting the control signals like the ISM band. The problem with this solution is that the control channel becomes highly crowded with the large number of cognitive users sending control messages. Hence at some point the performance of the network deteriorates due to channel saturation. However there is another solution like dynamically changing control channels. In this the control channel is 
allocated from one of the vacant channels and the channel keeps on changing with the activity of the primary users and according to the hopping sequence of the CR nodes. Following section gives a brief of some of the existing MAC protocols for CRNs. 

\subsection{MAC protocol for Centralised CRN}
Centralised CRNs have a central base station which manages the network activities, synchronisation and the coordinate the activities among nodes. Generally the base station is static and the CR users are at one hop distance from the base station. Hence the centralised CRNs allow an efficient coordination among the CR users in collecting network environment information.

\subsection{CSMA MAC for CRN}
\begin{figure}[!t]
\centering
\includegraphics[scale=0.4]{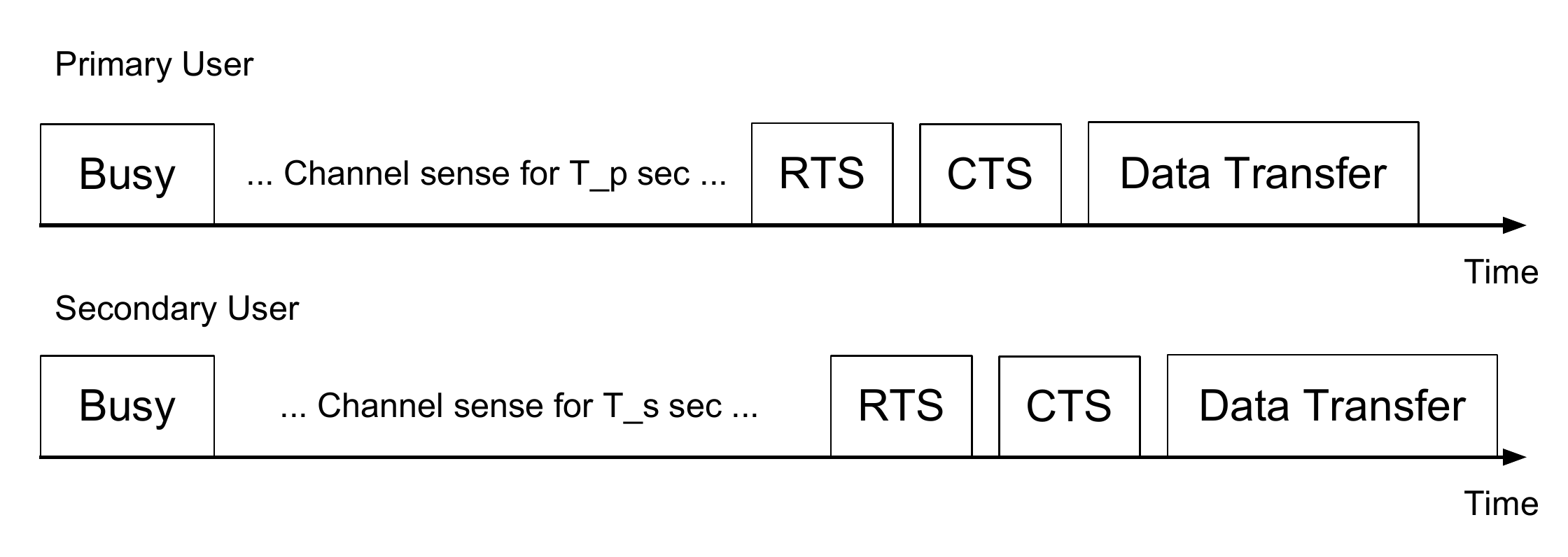} 
\caption{Illustration of the CSMA based protocol}
\label{fig21}
\end{figure}

The MAC protocol described in [10] is based on the Carrier Sense Multiple Access (CSMA) and uses one transceiver per CR user. The protocol proposed allows both the PU and the CR users to transmit simultaneously without affecting the PUs performance. The PU operate on classical CSMA protocol and undertake the carrier sensing for a time $\tau_p$. The CR users also work on the CSMA protocol and they sense the channel for time $\tau_s$. However the sensing time for PU is less than that of CR users ($\tau_p < \tau_s$) so as to give more priority to the primary users, as shown in figure \ref{fig21}. After the sensing phase the nodes send the request to send (RTS) and subsequently receive the clear to send (CTS). For each negotiation the CR users send only one data packet. This helps reduce the interference level of the primary users. The base station decides the transmission power depending upon the distance of the CR users from the base station and the noise power level.

\subsection{IEEE 802.22 MAC protocol}
The IEEE 802.22 is the Wireless Regional Area Network (WRAN) standard for the Cognitive Radio Networks [11]. It is proposed for rural broadband wireless access. This MAC protocol is for the Centralised CRNs. It proposes to use the TV white space 54-862 MHz with the range of the network lying between 17-33 Km. The base station manages all the spectrum sharing and access among all the CR users in its premise. In the downstream direction, time-division multiplexing is used, while in the upstream demand based TDMA assignment is utilised. IEEE 802.22 MAC employs a hierarchy based frame structure in which the superframes are of duration 160 ms and each superframe is divided into 16 frames of size 10 ms each. The IEEE 802.22 MAC can also be operated without a dedicated CCC. The SU scans for vacant TV spectrum on start-up. The base station meanwhile broadcast the superframes in the vacant channels. The SU choose one of the vacant channels and receive control information from the base station on that channel and 
begins its communication.  

\subsection{MAC protocol for distributed CRN}
In the distributed CRN there is no central base station to coordinate the activities of the CR users. The various protocols are discussed as follows.

\subsection{CREAM-MAC}
The Cognitive radio enabled multichannel channel MAC protocol (CREAM-MAC) [12] utilises a CCC for channel contention. Each CR user has sensors which sense for vacant channel. The CCC is either a dedicated CCC or one of the most unused licensed channels. The nodes contend on the CCC using the binary exponential backoff-based IEEE 802.11 with Distributed Coordination Function (DCF). Once a channel is reserved by a particular user all the nodes make sure not to contend for that channel. This removes any chance for collision. 

\subsection{DH-MAC}
The dynamically hopping control channel based MAC protocol is proposed in [13]. In cases where a dedicated CCC is unavailable the hopping based control channel could be successful. Each CR user keeps on changing its channel according to its channel sequence. Each CR user stays on a particular channel for a definite amount of time. Each SU transmits a beacon signal along with its channel hopping sequence on every vacant channel it switches to. Any user desiring to communicate with this user follows the channel hopping sequence of its destination node. Hence both the CR users follow the same sequence and send data packets. If the channel is unavailable due to primary user activity the CR users wait for switching over to next channel in the channel hopping sequence.\\
\indent The problem with the dynamically hopping control channel is that it introduces unnecessary delays in the communication because tracking of control channel is time consuming process.

\subsection{SYNC-MAC}
\begin{figure}[!t]
\centering
\includegraphics[scale=0.4]{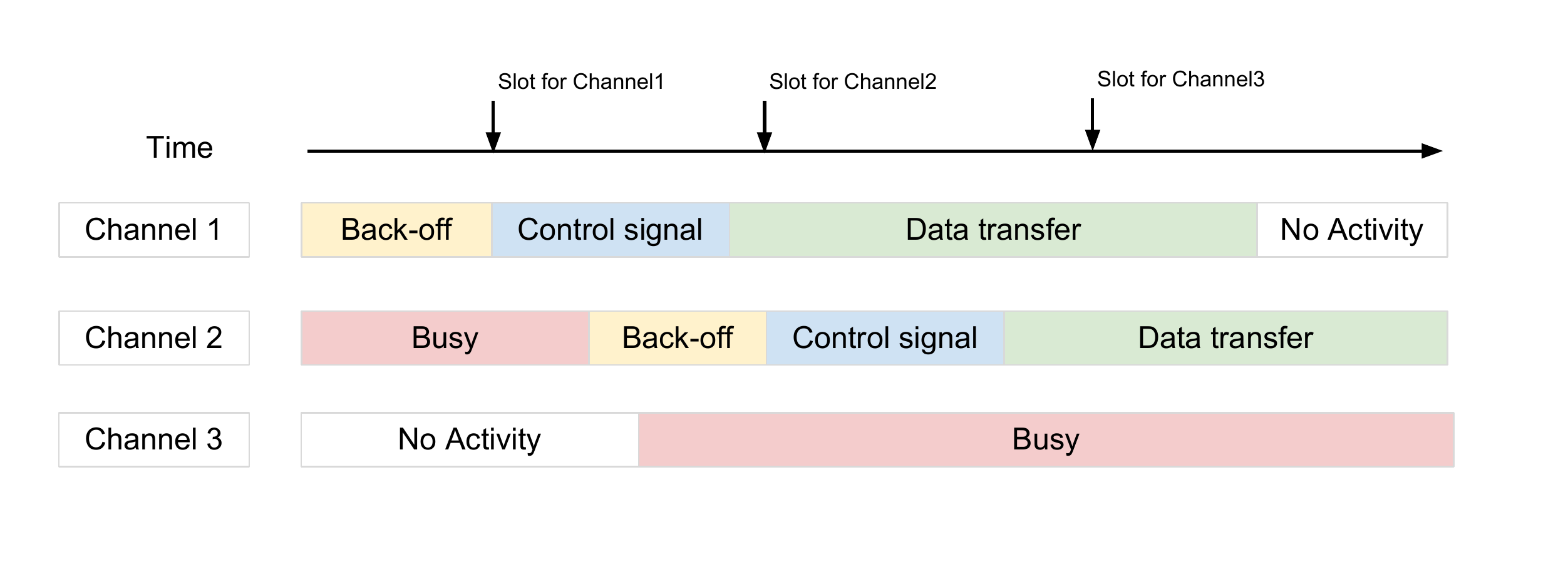} 
\caption{Three channels with control and data slots schedule as in SYNC-MAC}
\label{fig22}
\end{figure}

The synchronised medium access control (SYNC-MAC) proposed in [14] is a MAC protocol that does not either requires a CCC or a hopping based control channel for its operation. Hence eliminates any chance for control channel saturation or introducing unnecessary delays. Each node has two radios per node, one for listening control messages on the channel while the other is used for data transfer. The time is divided into time slots. Each user scan for vacant channels and listen to the control messages. A node desiring to send data sends a control packet to reserve for the channel. After successful contention it starts sending the data on this channel. The protocol has a dedicated radio for listening hence the hidden terminal problem is removed. Figure \ref{fig22} shows 3 channels with the control and data slots in their respective slots. The problem with the protocol is that the vacant channel is used only once hence the efficiency decreases.\\

\indent There are many MAC protocols for the Cognitive Radio Adhoc Networks that mainly employ TDMA mechanism either at control channel or at the data transmission. These protocols generally use the CCC for negotiating among themselves. Some of these are as follows.

\subsection{Cog-PRMA Protocol}
Cognitive packet reservation multiple access (Cog-PRMA) is a MAC protocol for the CR users which shares the channel with the TDMA primary user [15]. The channel access time is divided into frames which are further divided into time slots, which makes it closely resemble to TDMA. Whenever a CR node has a packet to send it contends and reserves the available time slot using the slotted aloha mechanism. There is a special sensing mini-slot at the starting of the time slot to sense whether the PU is transmitting in that slot or not. On detecting for an idle slot the CR users contend for the slot and reserve them for their transmission. The node releases the slot after it has sent its packet. After the release of the slot, the other CR users can send their packets using the same mechanism.

\subsection{Two level QoS support based MAC protocol}
\begin{figure}[!t]
\centering
\includegraphics[scale=0.4]{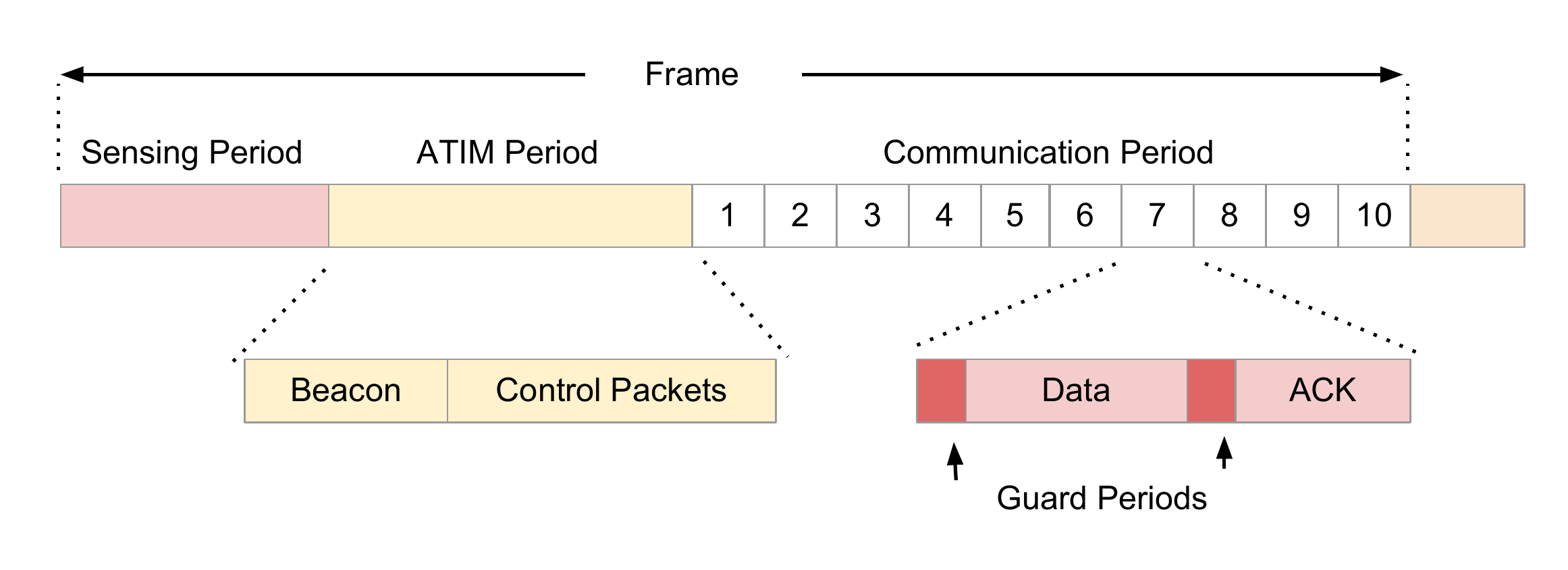} 
\caption{Frame structure for the MAC protocol proposed in [16]}
\label{fig23}
\end{figure}

In [16] the authors have proposed a MAC protocol which employs a CCC for carrying out slot contention and reservation. The system is divided into fixed length frames. Each frame is further divided into sensing period, ATIM period and communication period. Communication period is further divided into time slots for data transmissions. The nodes sense the channels in the sensing period for any PU activity and maintains a Channel occupancy list. In the ATIM period the nodes switch to the CCC to contend for the channel and time slot. The real-time traffic is given high priority by allowing the nodes to contend the slots first. Figure \ref{fig23} shows the frame structure employed in the proposed protocol. The protocol uses various interval times, for which a node waits before contention. This time is low for real time traffic and high for non-real time traffic.

\subsection{MAC protocol for TDMA based CR Networks}
The MAC protocol for TDMA based CRNs is proposed in [17]. The primary users are transmitting on the specific time slots hence some of the slots are left vacant. These are used by the CR nodes. This protocol consists of a CCC for communicating with the spectrum broker (responsible for channel allocation) and spectrum sensing. There are two transceivers per CR node, one tuned to the CCC and other is SDR which is capable of tuning to one of the primary channels. The control channel has three phases, allocation phase, request phase and sensing phase. In the allocation phase the spectrum broker sends a list of node id with the channel allocated to each. In the sensing phase the nodes sense their allocated channel. If the number of slots in the request phase is greater than the CR users then slotted aloha is followed otherwise TDMA is followed for slot request.
 
\subsection{MCR-MAC}
The Multi-channel cognitive radio MAC protocol (MCR-MAC) is proposed in [18]. This protocol applies TDMA and multi-channel to increase network throughput. CSMA/CA mechanism is employed to access the CCC for distributed coordination. This protocol solves the hidden incumbent node problem by using cooperative notification by the surrounding nodes.\\

\indent The use of common control channel causes a bottle neck in the system with high node density. There would be many nodes desiring to use CCC simultaneously hence the communication deteriorates. This is known as channel saturation problem. Most of the protocols proposed in the literature use CCC. The solution to CCC is proposed in DH-MAC where the author propose a hopping based control channel. This also has drawbacks like the delay in the system increases. To solve these problem we propose a novel MAC protocol in this thesis. The proposed protocol neither uses a dedicated CCC or a hopping based control channel. The vacant spaces in the spectrum act as control channel for some time and act as data slots for some time. The time is divided into time slots. The channels are scanned and whenever the channel is found vacant the secondary user transmissions are carried out. There are two phases namely control phase and data phase. In the control phase users send their requests in their respective time slot. 
While in the data phase the users send data packets in their allocated time slots.\\
\indent The proposed protocol maintains QoS by allowing users to request slots by sending their priorities in the control phase. The user with high priority is allocated the slot first. There are many protocols proposed in the literature to allow priority for real-time data traffic but it fails to suggest a solution for malicious user. A malicious user can act in such a way that it always gets access to the vacant channel, while creating loss for honest CR nodes. Our protocol uses the previous data slot allocation results to adjust the priorities of all the users. Hence even the low priority CR node also gets a fair chance to transmit.\\
\indent Power control is very important in case the nodes are battery operated.
The proposed protocol also discusses a mechanism to control transmission power
of the nodes and also allows node to go into sleep mode to save their power. The
following sections give a detailed description of proposed protocol along with
the simulation results.

%%%%%%%%%%%%%%%%%%%%%%%%%%%%%%%%%%%%%%%%%%%%%%%%%%%
%%%%%%%%%%%%%%%\input{paperch3}%%%%%%%%%%%%%%%%%%%%%%%%%%
%%%%%%%%%%%%%%%%%%%%%%%%%%%%%%%%%%%%%%%%%%%%%%%%%%%
% \section{Adaptive TDMA MAC Protocol}
\section{Proposed Protocol Design}
In this section we present the proposed MAC protocol named the Adpative TDMA MAC
protocol for Cognitive Radio Network (ATM-CRN). This protocol provides an energy
efficient solution for increasing the spectrum utilization. It also maintains
the Quality of Service(QoS) and guarantees a fair share of spectrum among the
Cognitive Radio Nodes.

\subsection{Protocol Description}
\begin{figure}[!t]
\centering
\includegraphics[scale=0.25]{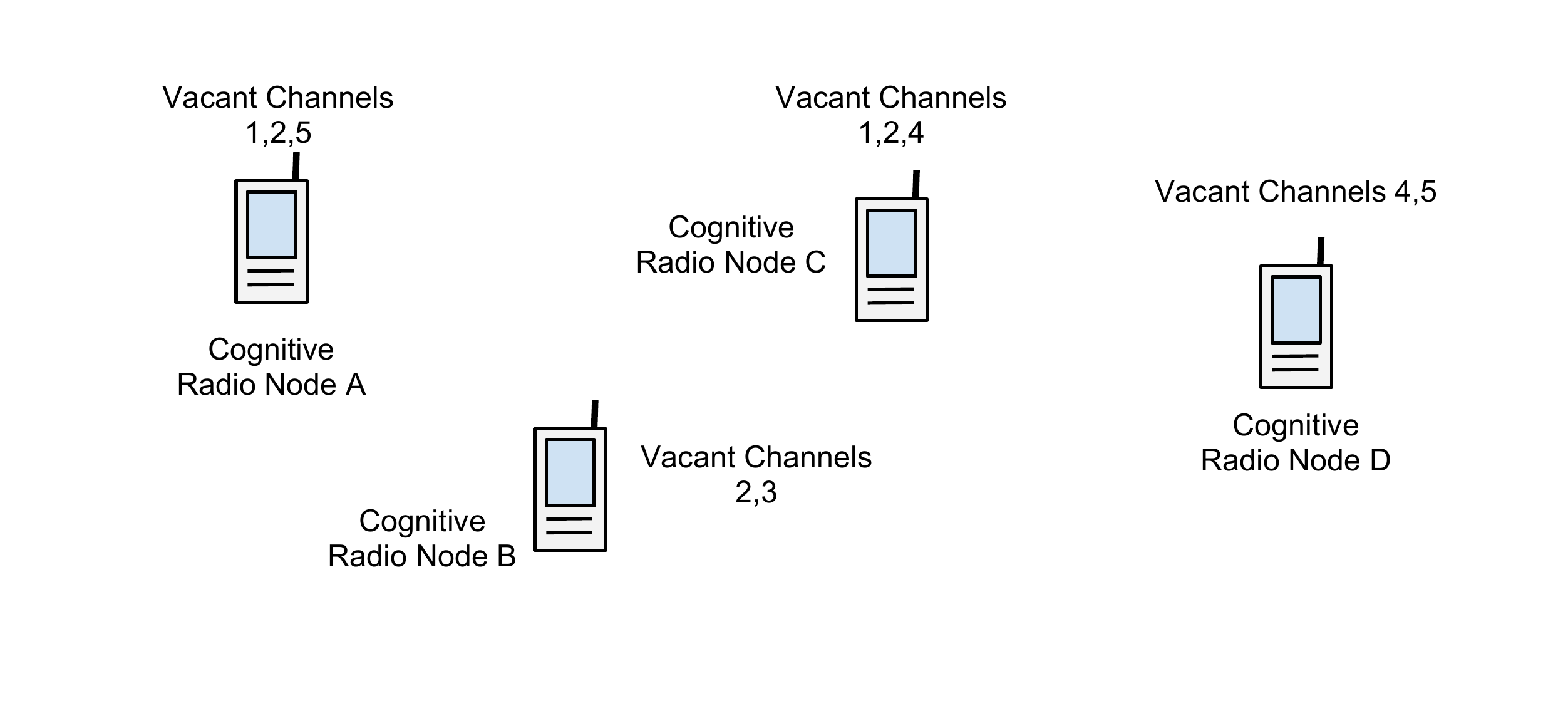} 
\caption{Cognitive Radio Nodes with vacant Channels}
\label{fig1}
\end{figure}

\begin{figure}[!t]
\centering
\includegraphics[scale=0.31]{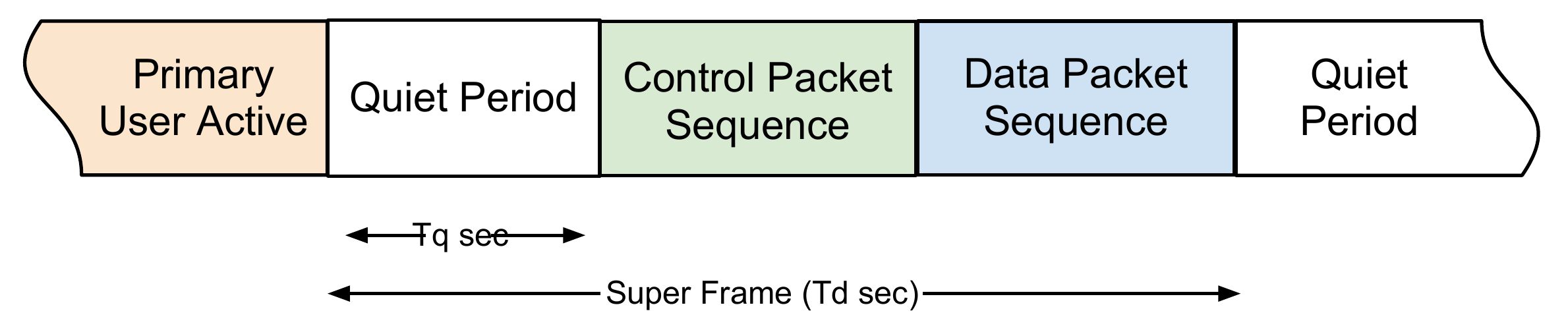} 
\caption{Packet Transmission Sequence}
\label{fig2}
\end{figure}

The ATM-MAC protocol for the Cognitive Radio Adhoc Networks is based on a non-Common Control Channel. The non-CCC for communication increases the reliability of such a network. Each CR Node scans for the channels which are vacant i.e. the channels in which there is no primary user activity. It maintains a list of these channels in its database and updates it at different times. Figure \ref{fig1} shows some Cognitive Radio nodes with the vacant channel detected by them. We have assumed that there is only one radio available per CR node. Each channel is divided into uniform time slots. The transmission is done over the allocated time slot. The contention for data slots is done in the specific time slots in the TDMA frame. Figure \ref{fig2} shows the outline of the packet sequence transmission. There are mainly two types of sequences, Control Packet Sequence and Data Packet Sequence.\\
\indent A superframe is the time starting from Quiet period to the end of data packet sequence.

\subsection{Quiet Period}
Quiet Period in Cognitive Radio is used to sense the channel for any primary user activity. The Quiet period is the time when all the nodes hear the spectrum for any PU activity. This is a minimum amount of time for which the CR users should sense the channel so that they can deterministically say that the primary users are not active in this channel or not. This depends on the type of communication carried out by the primary users. Once the primary users had subsided, the channel is sensed for $T_q$ seconds. This is the duration of Quiet period. The channel is sensed continuously at regular intervals for the return of the primary users. This interval's duration is taken as $T_d$ seconds.

\subsection{Control Packet Sequence}
\begin{figure}[!t]
\centering
\includegraphics[scale=0.34]{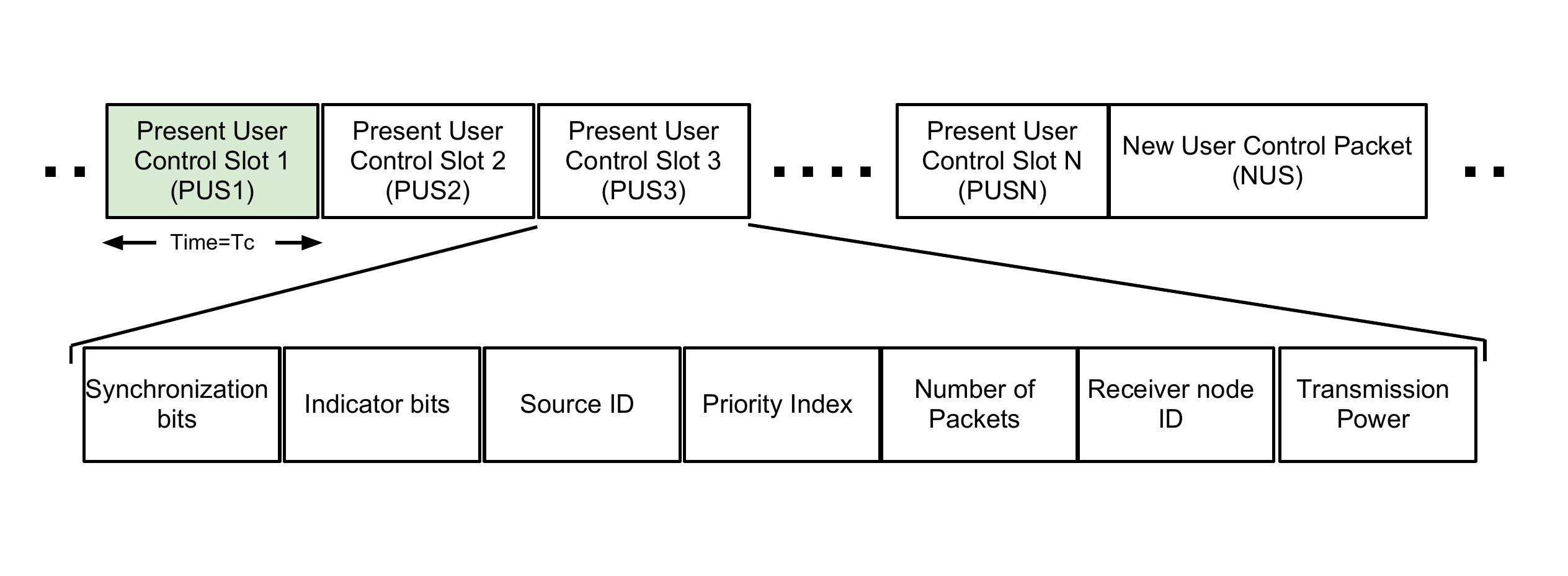} 
\caption{Control Packet Sequence}
\label{fig3}
\end{figure}

In the Control Packet Sequence each node broadcast a control packet over the network. Control packet include details like source ID, priority index, number of slots requested, destination node ID and transmission power of the control packet as shown in fig \ref{fig3}. Their description are as follows.\\

\begin{enumerate}
\item \textbf{Source ID} : The node ID of the CR Node sending Control Packet.
\item \textbf{Priority Index} :  Indicates the priority on which the network should process its request. It is calculated using the data type, number of packets in the queue and waiting time of the packets in the queue. The algorithm used for this calculation is Priority Index Calculator (PIC) described later. 
\item \textbf{Number of slots requested} : Number of slots that are required by the node for the transmission of packet. Each data slot is of fixed duration. According to the size of packet the node can calculate the number of slots required for transmission. 
\item \textbf{Receiving Node ID}: The destination ID to which the packet has to be sent (Could be to a particular user or can be a broadcast). In case the packet is to be broadcast this ID column has a special entry indicating broadcast.
\item \textbf{Transmission Power} : The transmission power at which the packet is being sent.
\end{enumerate} 

\indent The control sequence comprises of one new user slot (NUS) and several current users slots (CUS) as shown in figure \ref{fig3}. The CUS is reserved for sending of control packet by the current users in the channel. The NUS slot is reserved for the new users who want to take part in the communication for that channel. The new users send their control packets on NUS using CSMA/CD mechanism. The duration of this slot is twice the duration of the CUS. This is to ensure that if there are many users contending for this slot atleast there could be one CR user which could successfully send his control packet over this NUS.\\

\indent The number of control slots is adaptive to the number of active CR users in the network. All the users have the record of how many users are taking active part in the network and hence the number of CUS is also known to all the users. This mechanism allows for efficient use of the spectrum with no slot left vacant and also reduces the latency in the end-to-end delivery of packets. If a user in the CUS leaves the channel its CUS slot is healed by the successive user in the sequence by shifting their control slots in the next superframe. This concept would be further described in the working of the protocol.\\

\indent The transmission power of the packets are sent to all the nodes in the Control packet. All the nodes receive the Control packet and the power of the packet received. The nodes then compute the received and transmitted power to figure out how far the nodes are? Thus the Cognitive Radio nodes can adjust the power intelligently. This has two advantages, firstly, the power of the CR nodes are saved and secondly, the probability of interference with the primary users is also reduced.\\

\indent Along with the above five components the packets also contain the synchronisation bits (12 bits) for coherent detection at the receiver and the Indicator bits. Indicator bits are the bits added to indicate the type of packet. This is of 4 bits. Table \ref{tab1} indicates the different type of packets with their respective indicator bits. These two components are added to all the packets whether be control packet or data packet. \\

\begin{table}[h!]
\begin{center}
    \begin{tabular}{|l|l|}
    \hline
    Indicator Bits & Description \\ \hline
    0000 & Control Packet\\ \hline
    0001 & Common Data Packet\\ \hline
    0010 & Channel Control Packet\\ \hline
    1001 & Acknowledgement to Common data Packet\\ \hline
    1010 & Acknowledgement to Channel Control Packet\\ \hline
    \end{tabular}
\end{center}
\caption{Description of Indicator Bits}
\label{tab1}
\end{table}

\subsection{Data Packet Sequence}
\begin{figure}[!t]
\centering
\includegraphics[scale=0.29]{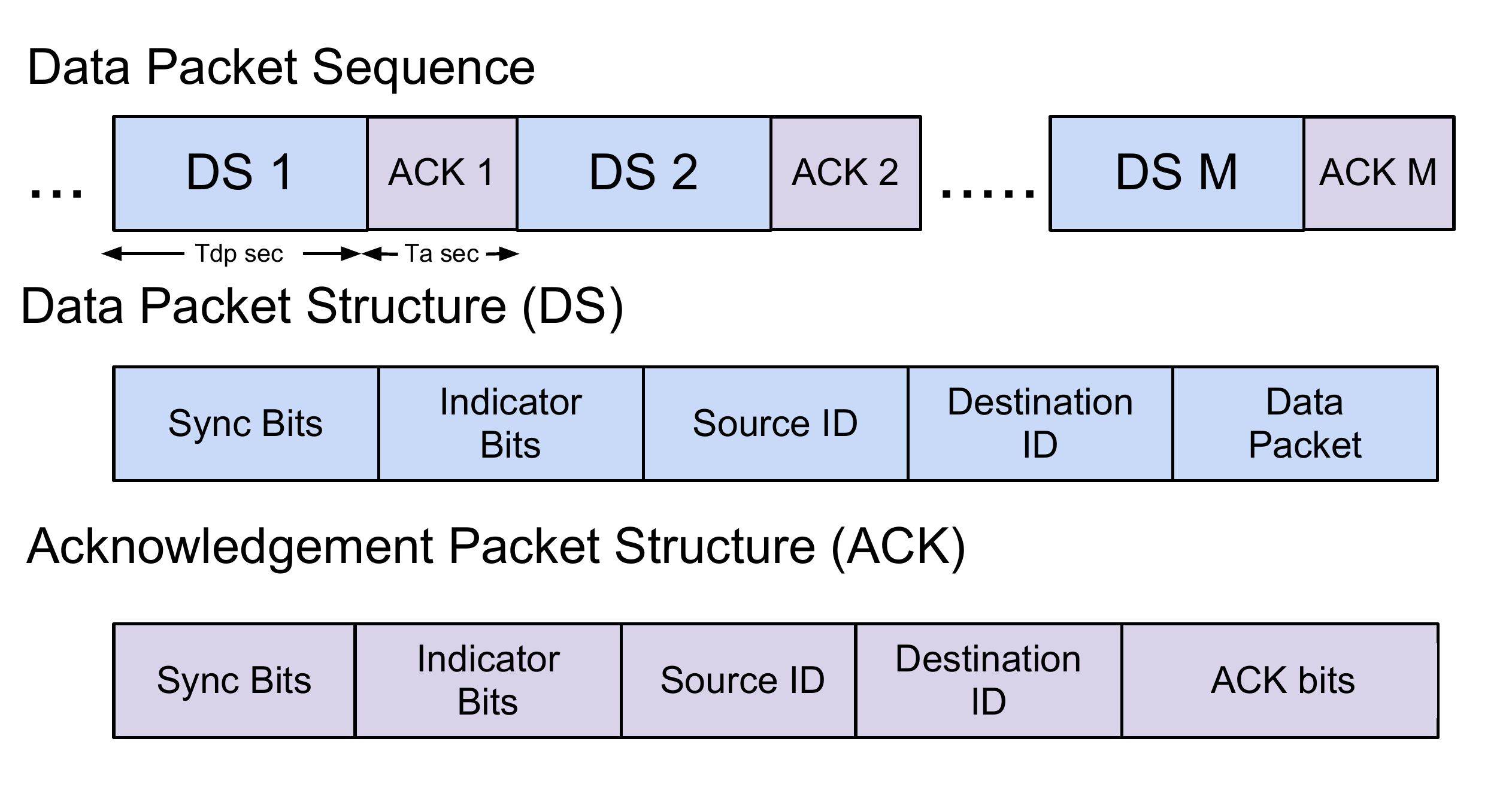} 
\caption{Data Packet Sequence}
\label{fig4}
\end{figure}

The data packet sequence consists of many data packet slots and acknowledgement
slots as shown in figure \ref{fig4}. Data packet slots are intended for
transmission of actual packets among the CR users in the Cognitive Radio
Network. The data packets can be distinguished from other packets by the
indicator bits. The structure and use of the special packets would be discussed
later in this section, but for now we would discuss the most general type of
data packet. The final destination of this packet is a user within this channel,
i.e. in this very hop the packet reaches its final destination. The Data Packet
consists of following details.
\begin{enumerate}
\item \textbf{Source ID :} The ID of the CR node transmitting the packet.
\item \textbf{Destination ID :}The ID of the CR node to whom the packet is to be transmitted. This CR node should be the one in this channel. This is the next hop MAC ID.
\item \textbf{Packet :} The Packet to be transmitted.
\end{enumerate}

\indent After the packet is received by the destination node, it sends an acknowledgement packet telling the sender that the packet is received successfully. The acknowledgement is send in the respective acknowledgement slot. This completes the first data packet transmission. The subsequent data packets are sent after the acknowledgement slot.\\
\indent The number of Data Packet Slots are adaptive. It depends on the number of data slots requested in the Control Packet Sequence. Each user stores the various parameters sent by the control packets from various CR users. Hence they keeps track of the number of data slots requested from all nodes along with the priority index. At the end of the Control packet sequence all the nodes use the Packet Scheduling Algorithm (details in section \ref{psa}) to calculate the distribution of the data slots among the CR users. The number of data slots are limited by the time $T_d$, the time of the superframe. If there are larger number of slots requested than it can accommodate in the given time, the network denies the transmission of some packets to meet the time-line. The time period of each data slot is fixed to $T_dp$ seconds and that of acknowledgement packet is kept as $T_a$ seconds. Hence M (highest number of data slots) can be decided using the equation 
\begin{center}
$ T_d=T_q+N\times T_c + M\times(T_dp + T_a) $
\end{center}
where $N$ is the number of users sending control packets.\\
\begin{table*}[t]
\begin{center}
    \begin{tabular}{ |l|p{1.8cm}|l|p{1.8cm}|}
    \hline
    Data Type & Sub-priority (DT) & Queue Length & Sub-priority (QL) \\ \hline
   Text data, File Transfer & 0 & 1-5 Packets & 0 \\ \hline
   Real-time voice and video & 1 & 1-10 Packets & 1 \\ \hline
   Control data Packets & 2 & 10-20 Packets & 2 \\ \hline
   Safety Critical data & 3 & $>$20 Packets & 3 \\ \hline    
    \end{tabular}
\end{center}
\begin{center}
    \begin{tabular}{ | l |  p{1.8cm} |}
    \hline
   Packet Delay & Sub-priority (PD) \\ \hline
   $<$2 superframes & 0\\ \hline
   2-5 superframes & 1\\ \hline
   5-10 superframes & 2\\ \hline
   $>$10 superframes & 3\\ \hline    
    \end{tabular}
\end{center}
\caption{Sub-priority index for each component}
\label{tab2}
\end{table*}
\subsection{Priority Index Calculator}
In the adhoc network environment, the various CR nodes could have different types of data to be transmitted. These data types could range from simple file transfers to real-time telephony and safety critical communications. These data types have their inherent properties and needs which gives rise to priorities. Some data types like voice calls, video calls etc. are to be transmitted with minimum latency while the data type like file transfers, text messaging etc. could be transferred with a tolerable end to end delay. As the spectrum availability in the CRN environment depends on the PU activity and is highly dynamic, the sufficient availability of spectrum to enable all the CR nodes to send their data is highly doubtful. Also it is recommended to give priority to real-time data transfers than the other off-line data, hence we introduce the concept of the Priority Index (PI). Priority index also ensures the fairness among the CR nodes, by giving equal opportunity to all the nodes to send their data.\\
\indent PI depends on the data type as well as on the number of packets in the queue and the waiting time of the packets in the queue. There are three main components to calculate the priority index for a CR node. Each priority component has a sub-priority index associated with it.
\begin{enumerate}
\item \textbf{Data Type (DT):} Sub-priority index is assigned according to the type of data. High to real-time data and low to delay tolerant data.
\item \textbf{Queue Length (QL):} Sub-priority index is assigned according to the number of packets in the queue, high to large number of data in the queue.
\item \textbf{Packet Delay (PD) :} Sub-priority index is assigned according to the delay in transmission of the first packet in the queue.
\end{enumerate}  
Table \ref{tab2} gives the different values of the components of the priority index. The priority is calculated using the formula.
\begin{center}
$ Priority Index = {3\times DT + QL + 3 \times PD} $
\end{center}
Higher priority index indicates urgent need by the CR nodes to send the data. Please note that the PI can range from 0 to 21.

\subsection{Packet Scheduling Algorithm}
\label{psa}
The Packet Scheduling algorithm (PSA) is an algorithm running on all CR nodes to calculate the scheduling of data slots to the CR users. The PSA takes input the priority number of all the CR nodes, the number of slots requested by the CR node and the past output of the PSA. The past outout of PSA refers to the list of CR users which were allocated a data slot in the last or previous superframe. By using these inputs the PSA calculates a net priority index (NPI). It allocates the high NPI node the first data slot and the low NPI nodes, the subsequent data slots. The PSA runs on all the CR nodes on the same set of data and then allocates the data slots to the CR nodes, hence there is no discrepancy in the sequence among the CR users.\\
\indent The past output of PSA ($PPSA_i$) of $i$th CR user is defined as equal to 5 if the data slot was not allocated in the last superframe else 0 if the data slot was allocated in the last superframe. The algorithm for the Packet Scheduling Algorithm is as follows.
\begin{enumerate}
\item Add PPSA to priority index of all CR users to form net priority(NP). $NP_i=PI_i+PPSA_i$
\item Arrange the list of CR Users according to the decreasing order of net priority index (NPI).
\item In case the net priorities are same for multiple users, sort them according to the decreasing order of $PPSA_i$. 
\item Starting with the CR node having highest net priority (NPI), assign the data slots to that CR User. The number of slots requested indicate the number of continuous data slots assigned.
\item The number of data slots are variable with a upper bound equal to the time between the Quiet periods. If there are less requests for data slots, the data slots allocated are less. If the requests are more then some of the nodes with low NPI are not allocated the slot. The maximum number of data slots that can be be allocated $M$ can be found using the equation,
\begin{center}
$ T_d\geq T_q+N\times T_c + M\times(T_dp + T_a) $
\end{center}
where $N$ is the number of users sending control packets.
\end{enumerate}

\subsection{Working of Proposed Protocol}
In this section, complete working of protocol is described in steps as follow.
\subsubsection{Initialization}
The CR nodes upon start-up senses the channels for white spaces (vacant bands). CR node selects a vacant channel at random to carry out the communication. The initialisation of communication starts with the transmission of control packet over the channel. The CR node having a packet to send, initiates the process by sending a control packet over the channel. This is send when there is no CR user activity in the network. This is first packet sent over the channel by any CR node. In the Control packet sequence this packet occupies the new user slot (NUS). On receiving this control packet by all the users in that particular channel, the other users also transmit their control packets. This is carried out by the following process.

\subsubsection{Entry of new users in the channel}
\begin{figure}[!h]
\centering
\includegraphics[scale=0.29]{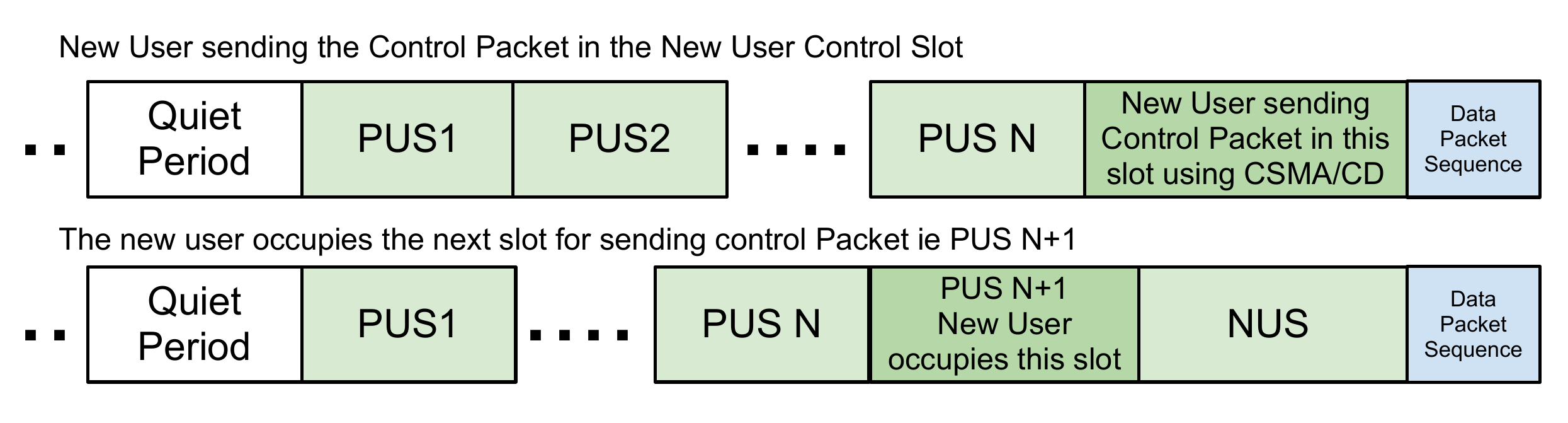} 
\caption{Control Slot occupied by New users in the Control Packet Sequence}
\label{fig6}
\end{figure}

Whenever a new node wants to communicate in a particular channel where the communication is already in progress (atleast one node is sending control packets), it has to send its control packet over the new user control packet slot (NUS). This is carried out after examining a super frame to know which all users are currently operating on that particular channel and when to transmit. The node synchronises using the beacon transmitted during the control packet transmissions. After successfully transmitting in the NUS, all the users present in that channel register him on this channel. In the next super frame this new CR node occupies the next vacant control slot in the control slot sequence. Hence the lenght of the control sequence increases by one. Likewise other nodes are added by first transmitting on the new user control slot and then occupying the next vacant control slot. Transmission in the new control slot (NUS) is carried out using CSMA/CD protocol. Figure \ref{fig6} indicates the transmission of 
control packets over the control packet sequence for the new users.\\

\indent As explained earlier the number of control slots are variable depending on the number of users participating in the channel. The upper limit is fixed so that there is a maximum number of user that can take part in communication at a time. This number depends on the difference between the Quiet periods of the primary users. It should be such that it does not causes harmful interference to the primary users. The formula
\begin{center}
$ T_d\geq T_q+N\times T_c $
\end{center}
gives the maximum number of users $N$ that can be accommodated in the channel at a given time. As the new user wants to enter a particular channel, it tracks the communication on that channel for one superframe. In this process the new CR usr gets intimated with the number of CR users operating on this channel. If this number is greater than $N$ then the new user leaves this channel and chooses some other channel.\\

\indent During the communication process it is possible that a node has no packet to transmit and wants to switch channel or get out of the network. This is done by the following process.

\subsubsection{Leaving the channel or communication}
\begin{figure}[!t]
\centering
\includegraphics[scale=0.29]{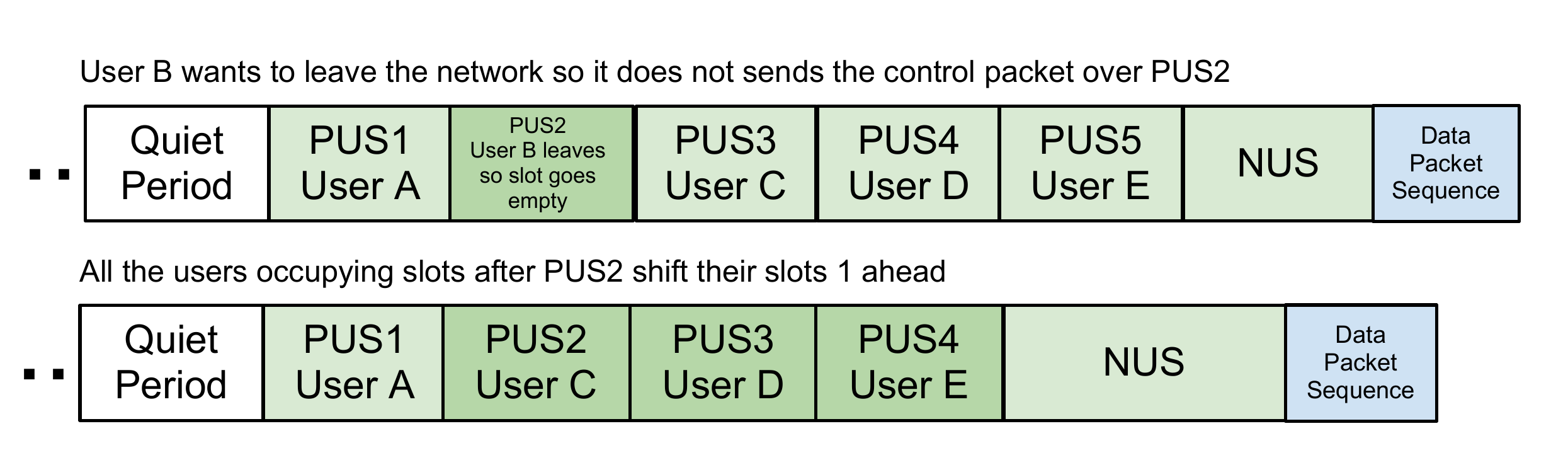} 
\caption{Control Slot distribution when a node leaves the network}
\label{fig7}
\end{figure}

Whenever a node wishes to leave a  particular channel, it just stops transmitting the control packet over the channel. All the other CR users not receiving the control packet by this CR user, assume that the node is absent. Hence they all update their control slot positions by shifting places as shown in figure \ref{fig7}. If there are large number of users present over a certain channel and all of them are not taking active part in communication, there is a mechanism proposed for the nodes to go in sleep mode. The users which have no packet to transmit can temporarily leave or pretend to leave the channel. This is done in a similar way by stopping the transmission of control packets. In the sleep mode the CR node does not transmit any packet but it receives all the control packets from other active users in the network. This is required because any packet intended for this CR node must be received by it. Also any control packet or network management packet must be received in order to adjust its functioning.
 Whenever they have to transmit they enter the channel using the mechanism proposed in the previous section.\\ 

\subsubsection{Transmission over data slots} 
After all the nodes have sent their control packets and received the control packets from the other nodes they would use the Packet Scheduling Algorithm to determine the order in which these nodes would send the data packets. According to the priority of the data packets and the number of packets from each user the algorithm gives the schedule for each node to send the data packet. Each node follows the same set of algorithm hence there is no chance of any ambiguity. After the Control packet sequence transmission the data packet sequence transmission is followed. After the successful reception of a data packet the receiver sends an Acknowledgement packet to complete the communication for a single data slot.\\
\indent After all the data packets are sent according to the Packet scheduling Algorithm, all the node sense for the channel for any primary users activity. This is Quiet period. After sensing the channel for a time $T_q$, if there is no primary user operation, it again sends a control packet. Likewise all the nodes in this particular channel send the control packets. The sequence they follow is the continuation of the previous control packet sequence. These control packets are followed by the data packets as stated earlier. \\
\begin{figure}[!t]
\centering
\includegraphics[scale=0.45]{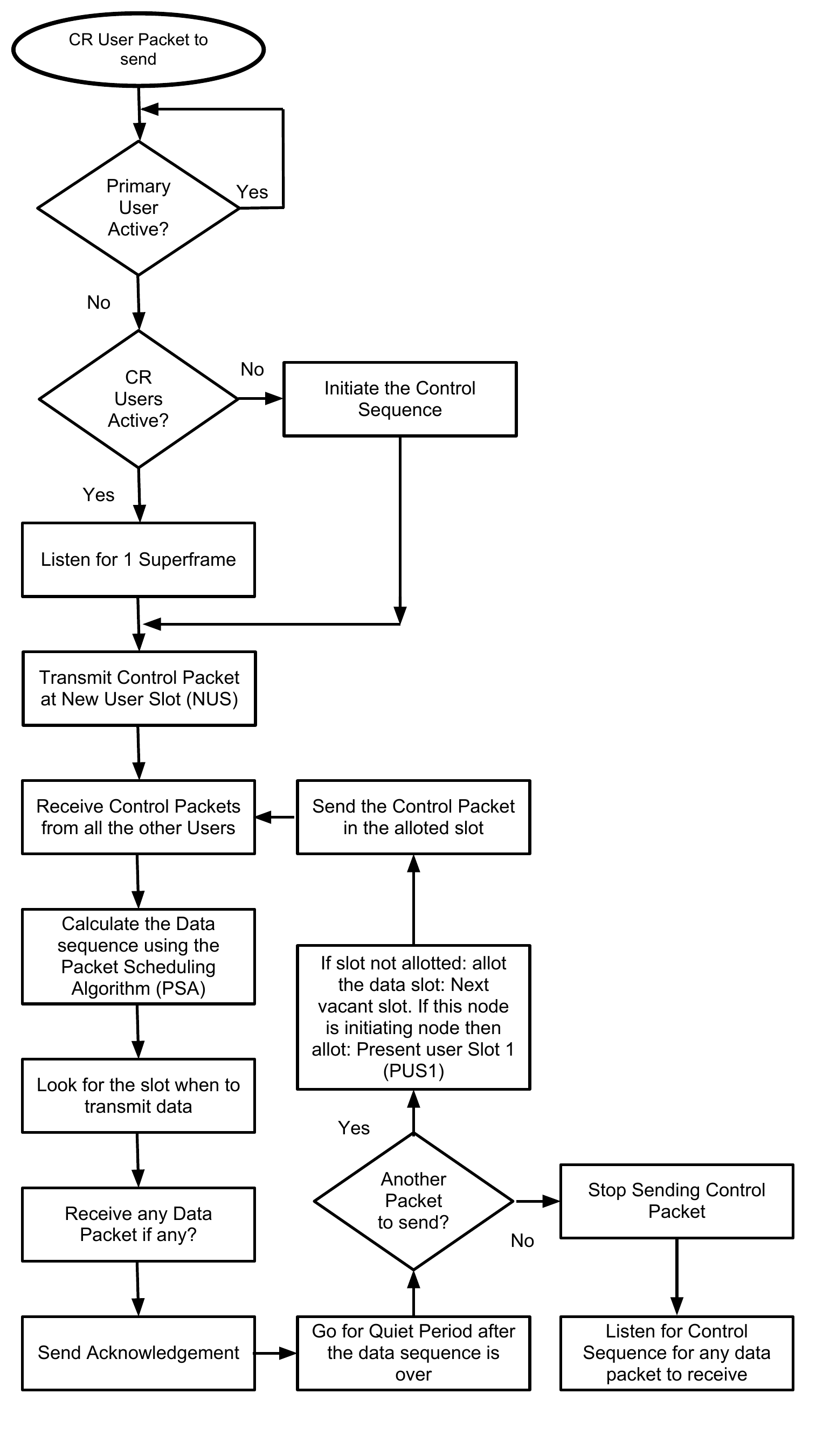} 
\caption{Illustration of the working of protocol using flowchart}
\label{fig5}
\end{figure}

% \subsubsection{Algorithm of the protocol}
The figure \ref{fig5} shows the complete flowchart of the step wise process
followed by each node in the network.
%  The algorithm is as follows.
%\begin{algorithm}

\subsubsection{Channel Selection by the nodes}
\begin{figure}[!t]
\centering
\includegraphics[scale=0.37]{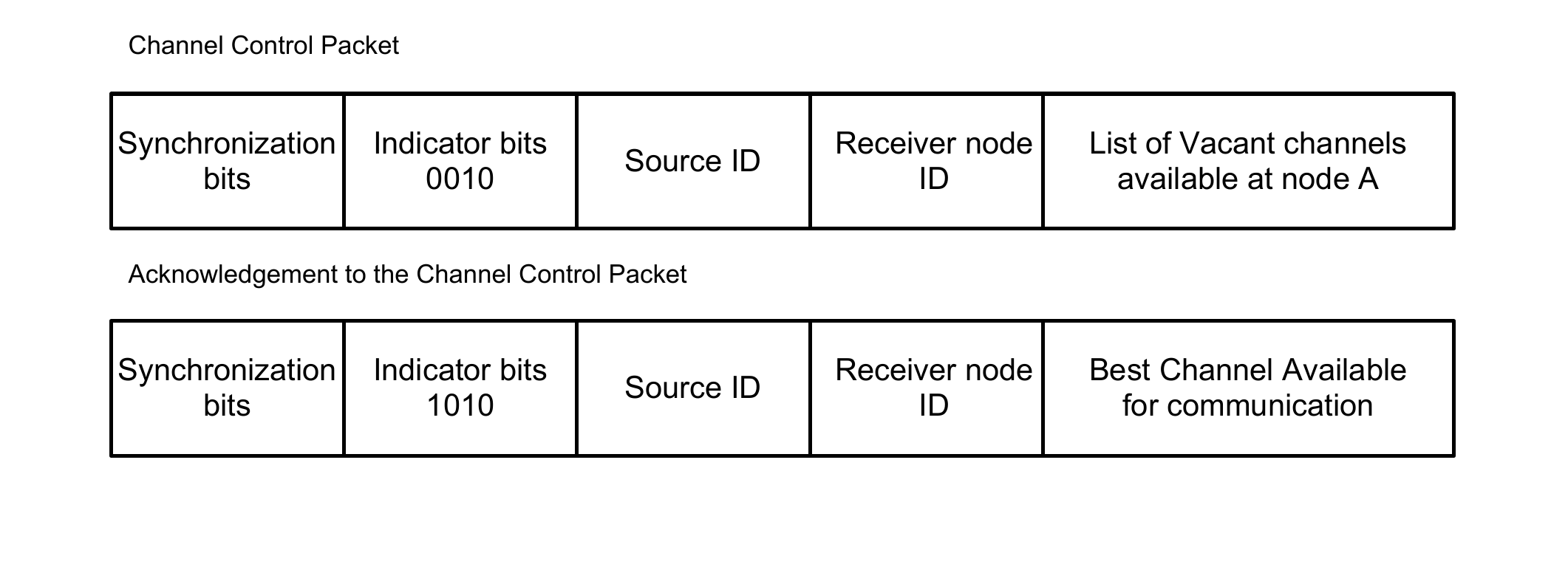} 
\caption{Structure of the channel control packet}
\label{fig8}
\end{figure}

In the Cognitive Radio Networks scenario there are cases when a particular vacant channel is highly crowded. Hence the problem that on which channel a particular user should operate arises. The choice of a channel for a particular CR User depends on the set of available channels for the receiver node and the number of CR users operating in that particular channel. For a single hop communication the set of available channels for the receiver and crowdedness of those channels play an important role. A special channel control packet is introduced to solve this problem. This packet is sent over the data transmission slots. This packet is a control data packet. The sender node sends a list of channels which it has detected as vacant to the receiver node on which they can communicate. The sender node examines all these channels and states which all channels are available with it for communication. Hence in response to the query the receiver node replies with the most suitable channel on which they can communicate. 
The sender node sends this channel ID along with the acknowledgement packet. The structure of the channel control packet is shown in figure \ref{fig8}. The algorithm is described next.
\begin{enumerate}
\item CR node A sends a Channel control packet to CR node B with the details of the channels available at its end.
\item CR node B receives the packet from node A and examines the channels mentioned in the packet.
\item CR Node B sends the channel most suitable to it in the acknowledgement packet to node A.
\item Node A after receiving the packet switches to the new channel.
\item Node B keeps on examining the superframe for the new channel. When it sees that Node A has entered the channel, it also enters the channel and begin communication.
\end{enumerate}
\indent This mechanism allows the channels to be uniformly occupied by all the nodes. If any channel has a high PU activity and CR nodes are not getting any opportunity to communicate then the nodes can switch to next vacant channel. The nodes wait for a wait time $T_w$ before switching to the next channel. In this switching each node upon expiry of wait timer switches to next channel and process continues till they do not find suitable channel for communication.
 
\subsubsection{An illustration of the proposed MAC protocol}
\begin{figure*}[!t]
\centering
\includegraphics[scale=0.34]{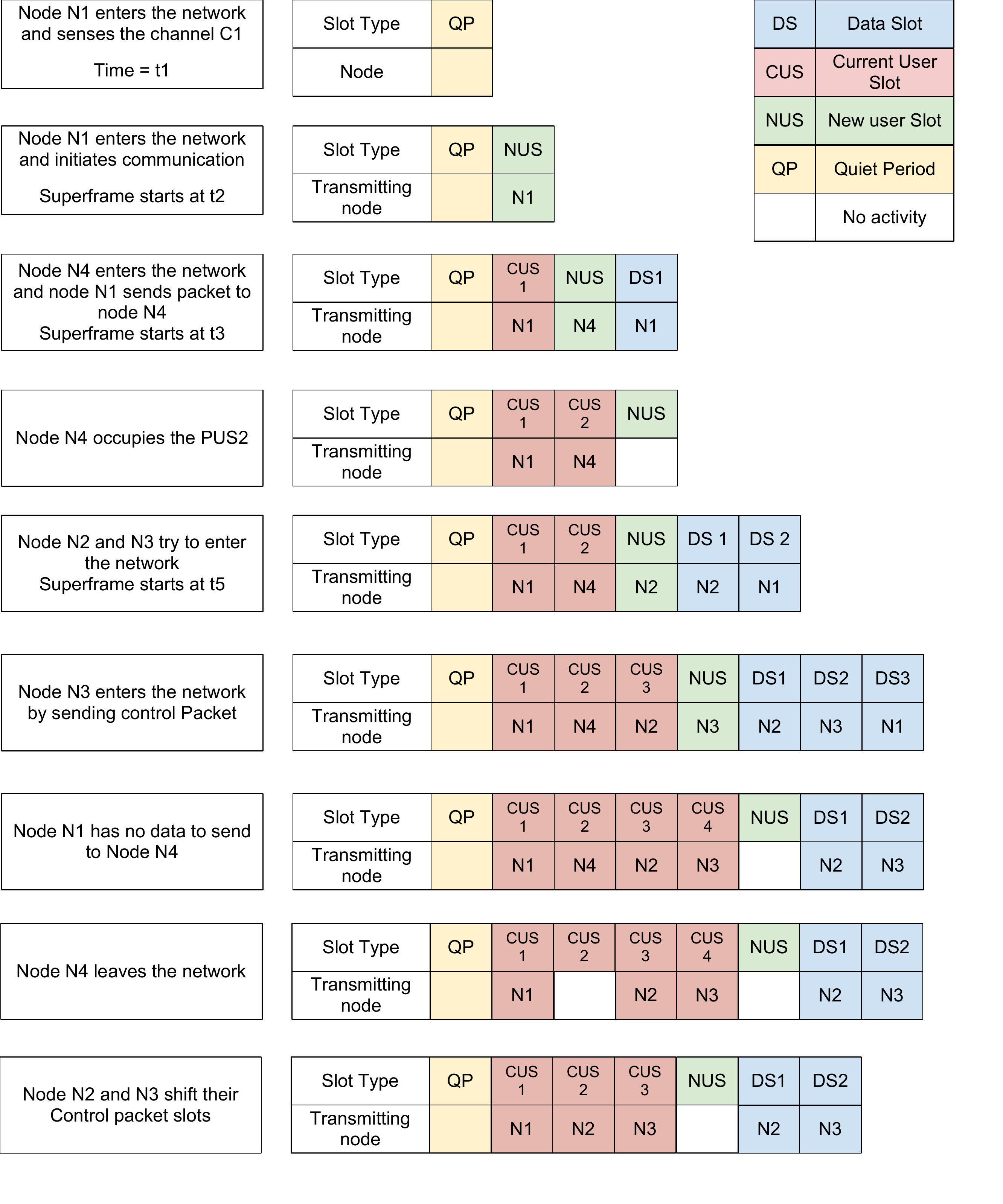} 
\caption{Illustration of nodes occupying the slots}
\label{fig9}
\end{figure*}

This section tries to illustrate with the help of an example the working of the MAC protocol for the Cognitive Radio Network.\\
\indent Let there be 6 CR nodes in the network at a single-hop distance. All the nodes can communicate to every other node in the network. All the nodes have detected channel $C_1$ as vacant and are tuned to this channel. At time $t_1$ CR node $N_1$ has a real-time video packet to transmit to CR node $N_4$. It senses the channel for $T_s$ time (Quiet Period) and then waits for another $T_s$ seconds for detecting any Cognitive Users' activity. On detecting that there is no activity on the channel, it sends a control packet over the channel at time $t_2 = t_1 + 2\times T_s$. The control packet has the information; Indicator bits=0000, Source ID = ID of $N_1$, Destination ID = ID of $ N_2$, Priority Index = 3, Number of Packets = 1, Transmission Power = $P_t$. All the nodes over the channel listens to the packet. At the next Super Frame, the node $N_1$ takes up the current user slot 1 (CUS 1) while New User slot(NUS) is kept vacant. The control packet sequence starts at time $t_3=t_2+t_cs+T_s$. At this time 
node $N_4$ registers itself with the network by sending a control packet over the NUS . The node $N_1$ sends its same control packet over the network on CUS 1. Both the node $N_1$ and $N_4$ execute the PSA algorithm at the end of the NUS to compute the data slot. At time $t_4=t_3+2\times t_cs$ the node $N_1$ sends a data packet to node $N_4$. In the next Super Frame the node $N_4$ occupies the CUS 2 for transmitting the Control packet, while the NUS is left vacant. Refer to the figure \ref{fig9} for the graphical illustration of the packet transfer over the network.\\
\indent At time $t_5$ node $N_2$ and $N_3$ have a safety critical packet to transmit to node $N_1$. In the next Super Frame the nodes $N_2$ and $N_3$ try to transmit the Control packet in NUS, using the CSMA/CD protocol. Lets suppose that node $N_2$ becomes successful in transmitting the packet. Hence it gets registered with the network, while node $N_3$ would try to send the control packet over the NUS in the next Super Frame. Similar procedure is carried out by the $N_2$ and $N_3$ to transmit the packets to $N_1$. Note that the priority index of node $N_2$ supersedes that of node $N_1$ hence node $N_2$ would occupy the first data slot in the data packet sequence.\\
\indent Lets suppose that the node $N_4$ wants to leave the network. From the next Super Frame it would not send any control packet over the network and the CUS 2 slot would go vacant. All the CR users de register $N_4$ from the list. On the next Super Frame all the CR users occupying the control slot after CUS 2 would shift their slot to compensate for the CUS 2 i.e. the node $N_2$ would now occupy CUS 2 and node $N_3$ would occupy CUS 3.

%%%%%%%%%%%%%%%%%%%%%%%%%%%%%%%%%%%%%%%%%%%%%%%%%%%
%%%%%%%%%%%%%%%%%%\input{paperch4}%%%%%%%%%%%%%%%%%%%%%%%
%%%%%%%%%%%%%%%%%%%%%%%%%%%%%%%%%%%%%%%%%%%%%%%%%%%
\section{Performance Evaluation}
In this section we will show the various simulation results for the proposed
protocol like delay and throughput parameters. We will also show the QoS
parameters for different types of data traffic. Using theoretical approach we
will derive the energy efficiency model for the same. The protocol has an
inbuilt structure to prevent bandwidth hogging by malicious users. This will
also be proved in the following sections.

\subsection{Throughput Analysis}\label{CCC}
\begin{figure}[!t]
\centering
\includegraphics[scale=0.35]{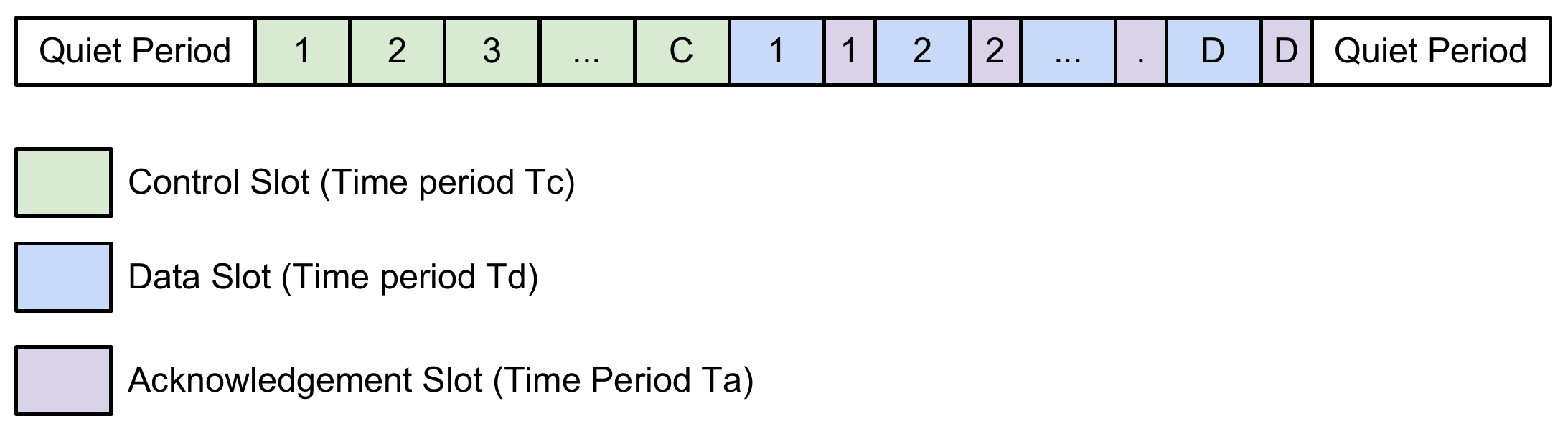} 
\caption{Slots distribution over a superframe}
\label{fig47}
\end{figure}

In this section we will derive the theoretical maximum throughput of the proposed MAC protocol. Figure \ref{fig47} shows the distribution of the slots as the control and data slots. Refer to table \ref{tab42} for the details of the various variables. \\
\indent The superframe time depends on the PU characteristics. Hence for $N$ number of users present the number of data slots available are
\begin{eqnarray}
T_s &=& N\times T_c + D \times T_d + T_q \nonumber \\
&&\textit{(Number of Control slots equal number of nodes)}\nonumber \\
D &=& \frac{T_s - T_q - N\times T_c}{T_d} 
\end{eqnarray}
Hence in 1 superframe $D$ number of data slots are available for transmission. If in one data slot $R$ bytes are transmitted then in 1 superframe number of bytes transmitted are 
\begin{equation}
\textit{Data transmitted in 1 superframe} = R\times D
\end{equation}
Hence the maximum throughput is 
\begin{eqnarray}
T&=&\frac{R\times D}{T_s} \nonumber \\
T&=&\frac{R(T_s - T_q - N\times T_c)}{T_s T_d} \label{eqn1}
\end{eqnarray}
The above equation gives the theoretical maximum throughput of the network.\\

\begin{table}
\begin{center}
\caption{description of various variables}
    \begin{tabular}{ | l | l |}
    \hline
   N & Number of CR nodes\\ \hline
   C & Number of Control slots per super frame\\ \hline
   D & Number of Data slots per super frame\\ \hline  
   $P_{tx}(R)$ & \parbox{2.5in}{\vspace{2pt}Power required to transmit packet to
a node at the distance R\vspace{2pt}}\\ \hline
   $P_{tx}(x)$ & \parbox{2.5in}{\vspace{2pt}Power required to transmit packet to
a node at the distance x\vspace{2pt}}\\ \hline
  $P_{rx}$ & Power required to receive a packet\\ \hline 
  	$T_c$ & Time period of a control slot\\ \hline 
  	$T_d$ & Time period of data slot\\ \hline
  	$T_a$ & Time period of acknowledgement slot\\ \hline
  	$T_q$ & Quiet period time\\ \hline 
    \end{tabular}
\end{center}
\label{tab42}
\end{table}

\begin{figure}[!t]
\centering
\includegraphics[scale=0.21]{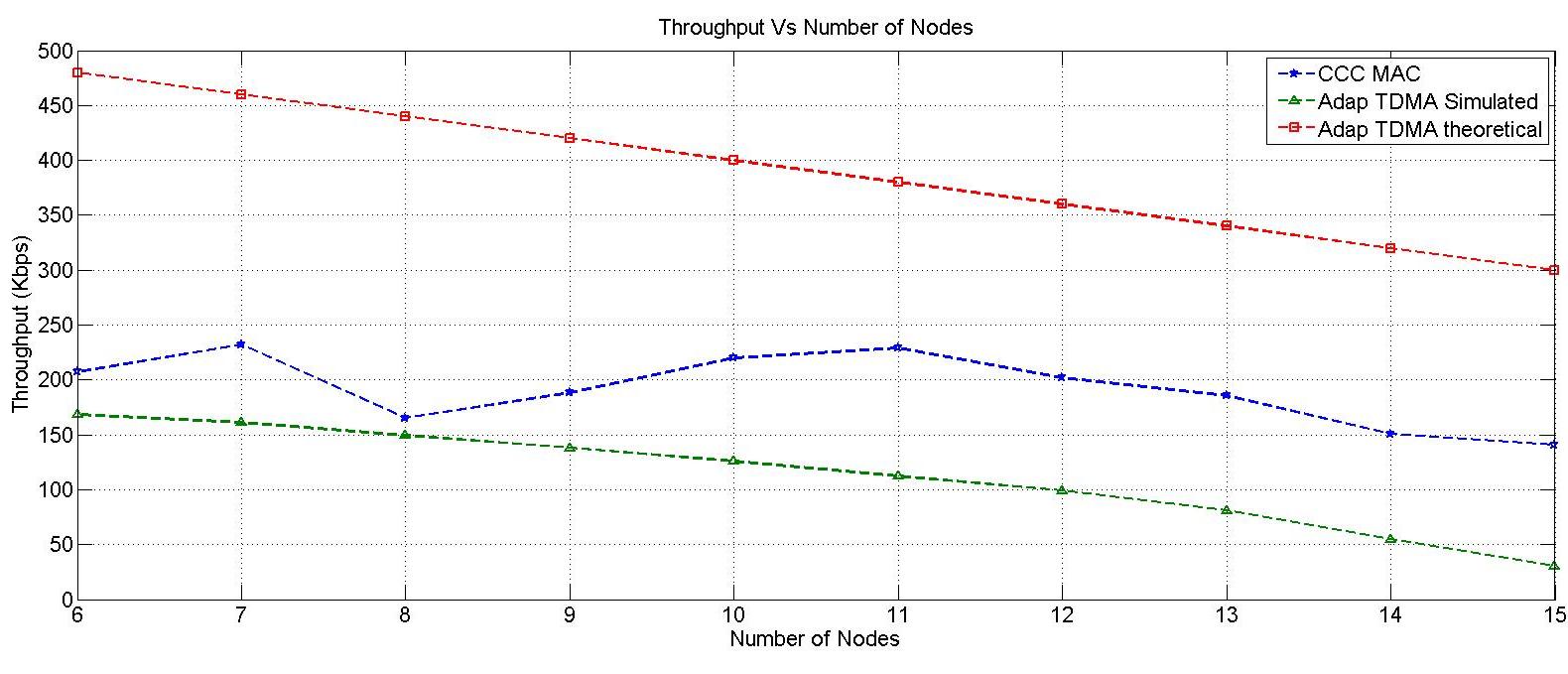} 
\caption{Graph between throughput and number of nodes}
\label{graph1}
\end{figure}

\begin{figure}[!t]
\centering
\includegraphics[scale=0.21]{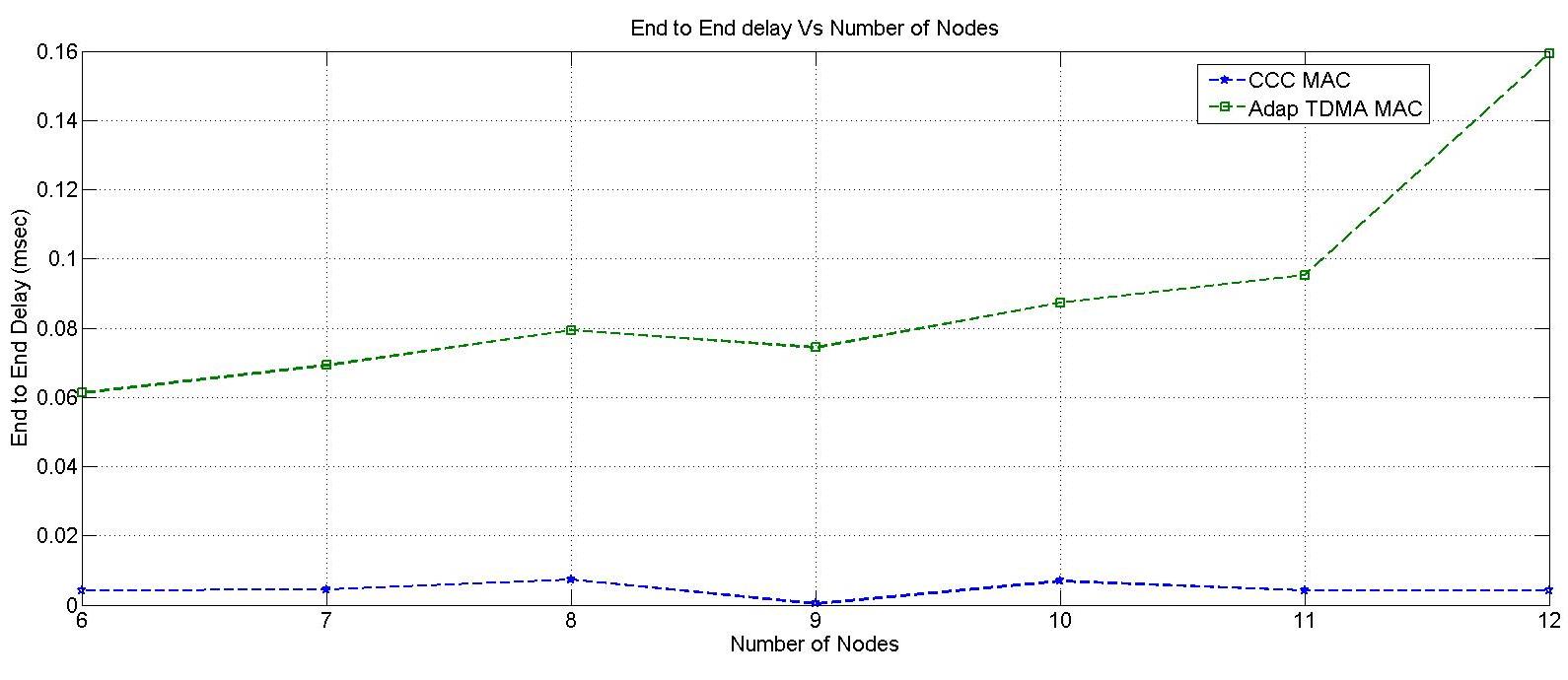} 
\caption{Graph between End-to-End delay and number of nodes}
\label{graph6}
\end{figure}

\indent We have simulated our protocol on ns-2 for evaluating its throughput performance. We have taken a single hop, single channel topology with each node having transmission range of 250m. The various parameters defined in the simulation is stated in table \ref{tab47}. Ad Hoc On Demand Distance Vector (AODV) routing protocol is taken for the routing of packets. The performance of our protocol is compared with the CCC MAC protocol.\\
\indent The CCC based MAC protocol is based on the Carrier Sense Multiple Access (CSMA) and uses a global control channel free from any primary user activity and always available to all the CR nodes. All the CR nodes sense the licensed channels for any PU activity and determine which all channels are free. Whenever a CR node has a packet to transmit, it sends out a request to send (RTS) packet over the control channel using the CSMA/CD mechanism. All the nodes in their idle state tune to CCC. On receiving the RTS packet, the destination node send a confirmation on clear to send (CTS) and they mutually agree on a channel. In the next step both the nodes switch to this channel and transmit their data. The various parameters used for simulation of CCC MAC is given in table \ref{tab47}. \\
\begin{table*}
\begin{center}
\caption{Values of various parameters for simulation in section \ref{CCC}}
    \begin{tabular}{ | l | l | l | l |}
    \hline
  \multicolumn{2}{|c|}{Adaptive TDMA MAC} & \multicolumn{2}{|c|}{CCC MAC protocol}\\ \hline
   Nodes & 6-12 & Nodes & 6-12\\ \hline
   Channel & 1 & Channels & 2 \\ \hline   
  	$T_c$ & 1.0 msec & Channel Bandwidth & 1Mbps\\ \hline 
  	$T_d$ & 1.0 msec & Packet size & 1000 Bytes\\ \hline
  	$T_a$ & 0.5 msec & Packet Frequency & 5 ms\\ \hline
Flow 1 & Node 1 to 2 (PI=20) & Flow 1 & Node 1 to 2 \\ \hline
Routing protocol & AODV & Simulation Time& 10sec \\ \hline
Source & CBR & Queue length	& 100 \\ \hline
Superframe Time Period & 60 msec & PU Activity & None \\ \hline
    \end{tabular}
\end{center}
\label{tab47}
\end{table*}
\indent The graph between throughput vs number of nodes is plotted for the theoretical analysis of protocol from equation (\ref{eqn1}), the simulation of our MAC protocol and the simulation of CCC MAC protocol over ns-2. The graph in figure \ref{graph1} shows that the throughput of our protocol decreases as the number of nodes increase. This is because more control packets are sent as the number of nodes increase, and hence number of data slots decrease for the same superframe time period. Also we can see that the throughput of the CCC MAC protocol higher than our proposed MAC protocol. This is because CCC MAC uses two channel, one as a dedicated control channel and other is the licensed data channel and our adaptive TDMA MAC protocol uses only one channel. Hence the CCC MAC has an advantage over our protocol in terms of the availability of a common channel.\\
\indent The graph between average end to end delay vs number of nodes is shown in figure \ref{graph6}. It is clearly seen from the graph that the curve for CCC MAC lies below the curve of Adap TDMA MAC. This is because we have used two channels (one control channel and one data channel) for simulating CCC MAC. Also our protocol has two phases one control packet transmission phase and another data packet transmission phase, hence there is a delay created when a packet arrives at the node during the control sequence phase. Hence the delay of our protocol is higher.

\subsection{Protocol sensitivity to various parameters}
\begin{figure}[!t]
\centering
\includegraphics[scale=0.21]{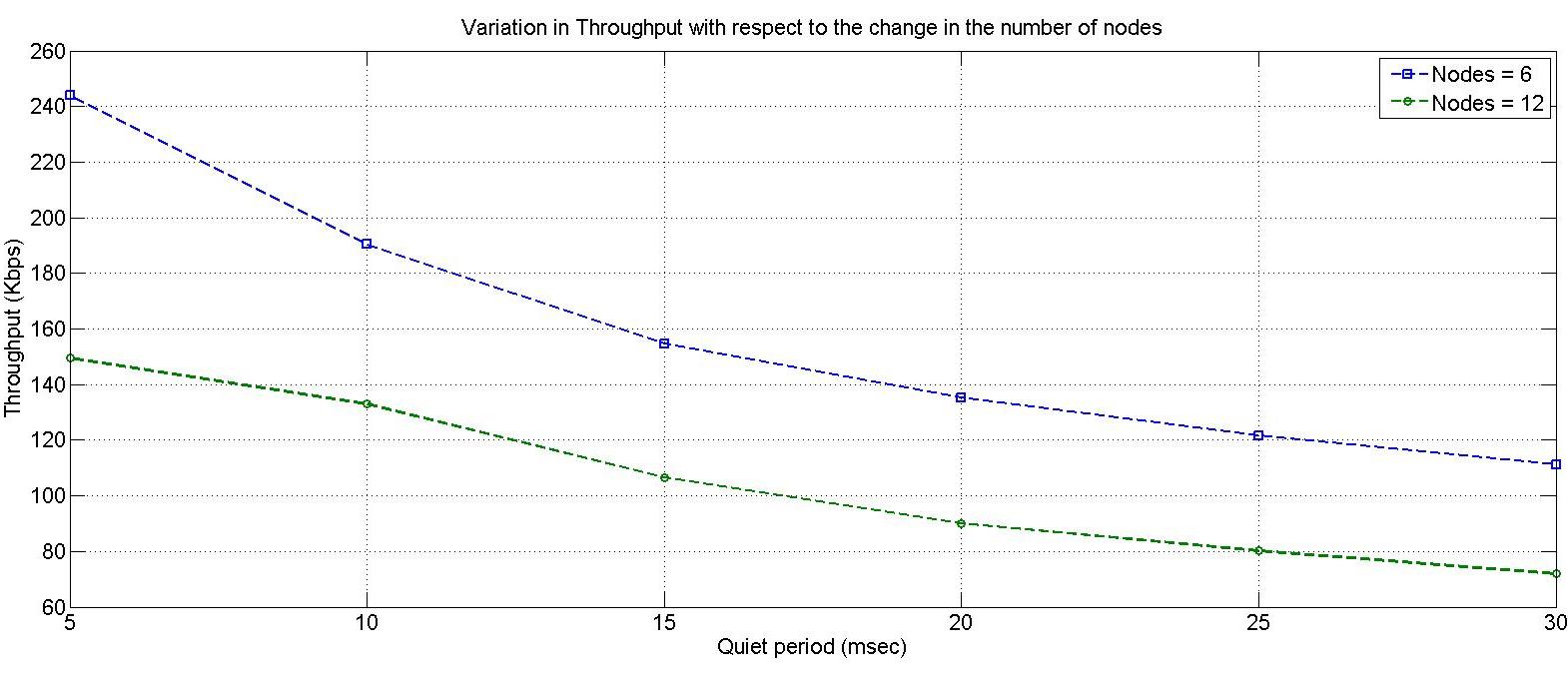} 
\caption{Graph indicating the variation in Throughput with respect to change in duration of Quiet period}
\label{graph2}
\end{figure}

\begin{figure}[!t]
\centering
\includegraphics[scale=0.21]{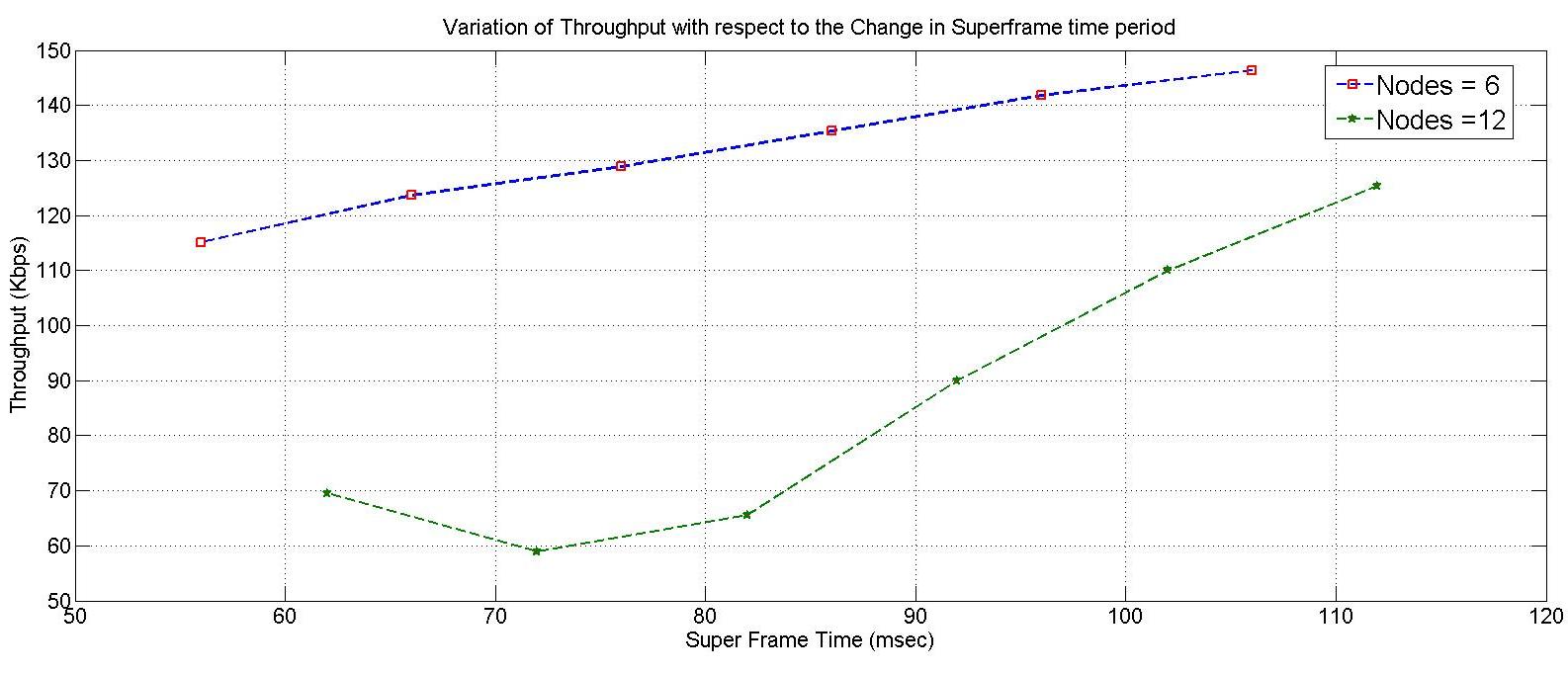} 
\caption{Graph indicating the variation in Throughput with respect to the change in length of superframe}
\label{graph3}
\end{figure}

\begin{figure}[!t]
\centering
\includegraphics[scale=0.21]{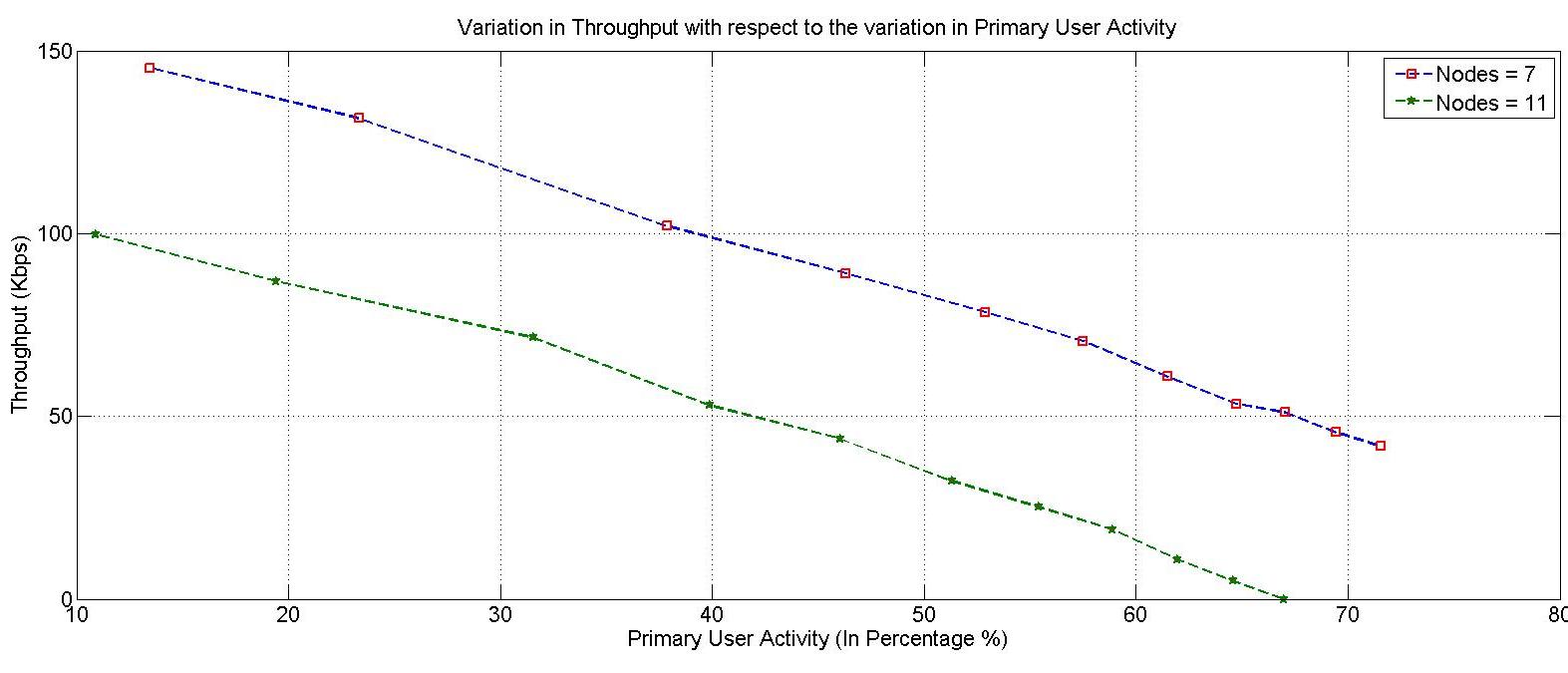} 
\caption{Graph indicating the variation in Throughput with respect to the variation in Primary User activity}
\label{graph4}
\end{figure}

\begin{figure}[!t]
\centering
\includegraphics[scale=0.21]{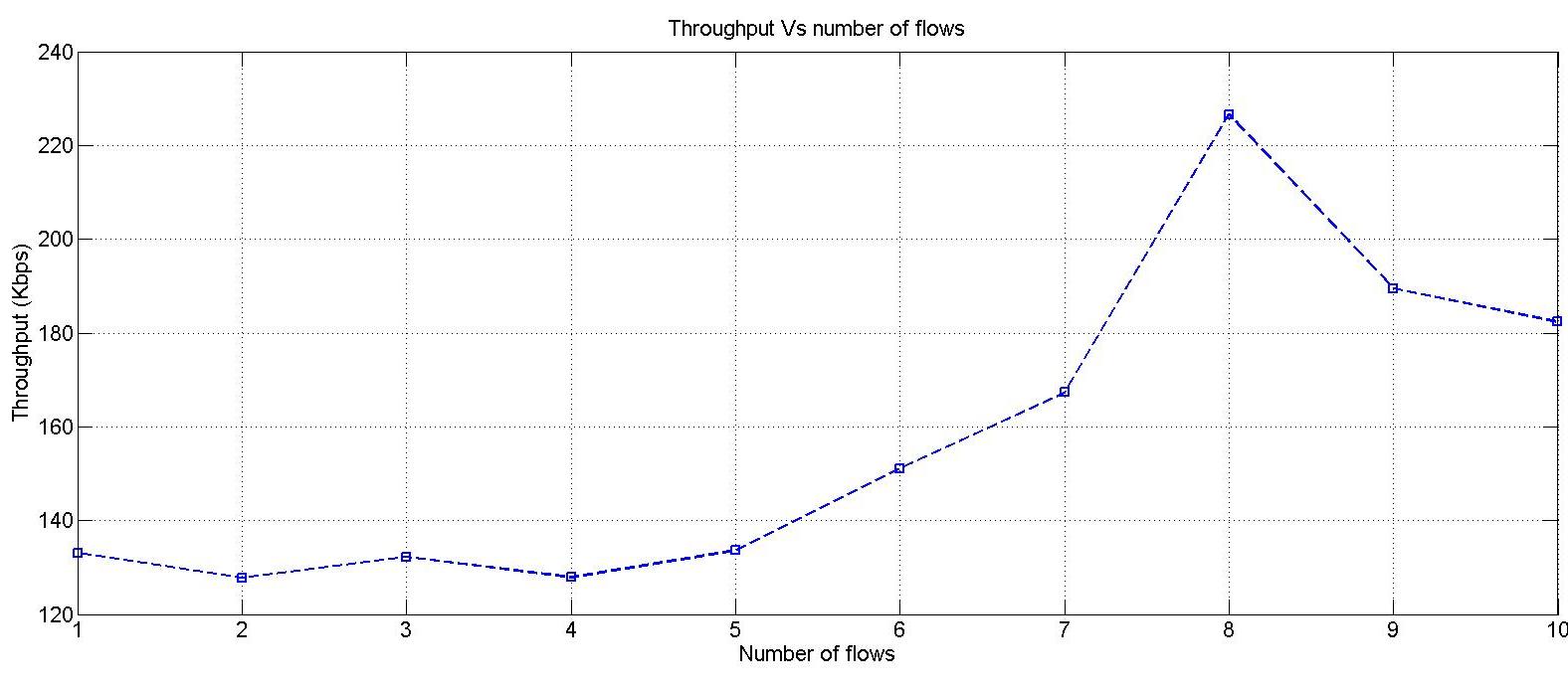} 
\caption{Graph indicating the variation in Throughput with respect to the variation in number of flows}
\label{graph7}
\end{figure}

\indent In this section we analyse the various parameters of the protocol influencing its performance.
We have analysed the throughput of our protocol by tuning the parameters like duration of quiet period, length of superframe and variation primary user activities.\\
\indent The graph \ref{graph2} shows the variation of throughput in case of varying the quiet period. The quiet period depends on the method employed for detecting the PU activity. Hence we vary the time of quiet period from 10ms to 30ms and simulate the protocol for 6 node network. During the quiet period we are not exploiting the spectrum, hence with the increase in quiet period the throughput decreases as seen from the graph.\\
\indent The graph \ref{graph3} shows the effect on throughput on the variation of superframe duration. The frequency of quiet period depends on the nature of primary user communication. This time is used for our communication which includes control and data packet transmissions. The control slots are fixed by the number of nodes participating in the network while the remaining time is utilised as the data slots. This gives a limit on the maximum number of data slots. We vary the superframe duration from ms to ms and record the throughput of the network under different conditions. The graph clearly shows that as the length of superframe decreases the utilisation of the spectrum decreases. The transmission of control packet acts as an overhead for the data sent in the superframe. Hence with less data slots, less data is send for the same control slots. Hence the overall throughput decreases as shown in the graph.\\
\indent This problem can be mitigated by releasing the limit on the number of data slots and introducing the quiet period in between the data slots. There could be quiet periods introduced in between five successive data slots to sense for PU activity. If the PU is detected the CR communication stops. The communication is resumed from where it left when the PU activity ceases.\\
\indent Next we show the throughput performance of our protocol for the variation in primary user activity in figure \ref{graph4}. We vary the PU activity and record the throughput for various scenarios. It is clearly seen from the graph that the throughput decreases with the increasing PU activity. 
\indent We also show the effect on throughput of the network with the variation in number of flows. Figure \ref{graph7} shows the throughput vs number of flows in the network of 12 nodes. All the other parameters are taken from table \ref{tab47}. We can see that as the number of flows increases the throughput of our system also increases. But after a certain limit as from the graph, due the throughput comes to a saturation level dictated by equation \ref{eqn1}. This is the maximum limit on throughput.

\subsection{Energy Efficiency}
\label{enr}
As discussed in the description of protocol there are slots in which the CR node has neither to send or receive packets. Hence the nodes can go off to sleep mode in these cases the power could be saved.\\
\indent Also in control packet we had included power of the packet transmitted. This information is used by all CR nodes to intelligently adapt their transmission power. The power of the received packet can easily be obtained from the physical layer. The transmitted power and received power of the control packet sent by a node is known to all other node at various places. Let received power of packet be $P_r$ and transmitted power of the packet be $P_t$. Hence according to the Frii's free space propagation model,
\begin{eqnarray}
P_r &=& \frac{P_t G_t G_r \lambda^2}{4 \pi^2 d^2 L}\\ 
P_r &\propto& \frac{1}{d^2}
\end{eqnarray}
where, 
\begin{itemize}
\item $\lambda$ is the signal wavelength (in metres).
\item $G_t$ is gain of transmitter antenna.
\item $G_r$ is gain of receiver antenna.
\item $d$ is the distance between the transmitting and receiving node (in metres).
\item $L$ is loss factor.
\end{itemize}

\indent Now we derive a mathematical formula for the amount of power saved by the nodes when we use our energy efficiency methods, i.e. frequent sleep mode of nodes and varying transmission power of the packets transmitted.\\
\indent We assume that the nodes are spread in a geographical area of radius $R$ meters. We will compute the energy consumption of the central node in the network in the case of uniform distribution of traffic among the users. We also assume that there is no primary user activity. The description of various variables in shown in table \ref{tab42}.

\indent First, we will derive the power consumption for the case when there is no mechanism of nodes going into sleep mode and there is no transmission power control among the nodes. We call this power $P_{wpc}$ (Power of node without power control). \\
Time period of 1 superframe \begin{equation} T_s = T_q + C \times T_c + D \times T_d + D \times T_a \end{equation}
Number of superframes per second \begin{equation} \xi = \frac{1}{T_s} \end{equation}
We have assumed that the traffic is uniformly distributed among the N nodes. Hence number of packet transmitted by each node in 1 superframe and packet received by each node in 1 superframe is \begin{equation} \lambda = \lfloor \frac{D}{N} \rfloor \end{equation}
Lets calculate energy consumption in 1 superframe. This energy consumption includes one control packet transmitted and C-1 control packets received. And there are $\lambda$ data packets transmitted, $\lambda$ data packets received and $\lambda$ acknowledgement packet transmitted and received. Note that the transmission power is taken as $P_{tx}(R)$ as the nodes have no capability to adjust transmit power hence they transmit the packet with maximum power.
\begin{equation}
\begin{split}
 E_{wpc} = P_{tx}(R) \times (T_c + \lambda T_d + \lambda T_a) + P_{rx} \times
[ (C-1)T_c + \\ (D-\lambda)T_d + 
(D-\lambda) T_a] 
\end{split}
\end{equation}
Hence power consumption is \begin{equation} P_{wpc} = E_{wpc} \times \xi \end{equation}
Now we calculate the power consumption of our protocol with transmission power
control mechanism and sleep off mechanism enabled. The energy consumption for 1
superframe $E_{pc}$ is 
\begin{equation} 
\begin{split}
E_{pc} = P_{tx}(R) \times T_c + P_{tx}(x)\times(\lambda T_d +
\lambda T_a) +\\ P_{rx}\times[(C-1)T_c + \lambda T_d + \lambda T_a]
\end{split} 
\end{equation}
Hence power consumption is \begin{equation} P_{pc} = E_{pc} \times \xi \end{equation}
The expected value of $P_{pc}$ is
\begin{figure*}[t]
% \begin{equation}
% \setcounter{MYtempeqncnt}
\begin{eqnarray} 
E[P_{pc}] &=& \xi E[E_{pc}] \nonumber \\
			&=& \xi E[P_{tx}(R) \times T_c + P_{tx}(x)\times(\lambda T_d + \lambda T_a) + P_{rx}\times[(C-1)T_c + \lambda T_d + \lambda T_a]] \nonumber \\ 
			&=& \xi [P_{tx}(R) \times T_c + P_{rx}\times[(C-1)T_c + \lambda T_d + \lambda T_a]] + \xi E[P_{tx}(x)\times(\lambda T_d + \lambda T_a)]  \nonumber \\
			&=& \xi [P_{tx}(R) \times T_c + P_{rx}\times[(C-1)T_c + \lambda T_d + \lambda T_a]] + \xi (\lambda T_d + \lambda T_a) E[P_{tx}(x)] \nonumber
\end{eqnarray}
% \end{equation}
\hrulefill
% The spacer can be tweaked to stop underfull vboxes.
\vspace*{4pt}

\end{figure*}
Now we have to calculate $E[P_{tx}(x)] $. The nodes are uniformly distributed over a sphere of radius R as shown in figure \ref{fig41}. From equation (4.1) we can get the relation between $P_{tx}(R)$ and $P_{tx}(x)$. 
\begin{eqnarray}
P_r &=& \frac{P_{tx}(R) G_t G_r \lambda^2}{4 \pi^2 R^2 L} \nonumber \\
P_r &=& \frac{P_{tx}(x) G_t G_r \lambda^2}{4 \pi^2 x^2 L} \nonumber
\end{eqnarray}
We require a constant power at the receiver so equating the above two equations, we get
\begin{eqnarray}
\frac{P_{tx}(R)}{P_{tx}(x)} &=& \frac{R^2}{x^2} \nonumber \\
P_{tx}(x) &=& \frac{x^2 P_{tx}(R)}{R^2} \nonumber \\
E[P_{tx}(x)] &=& \frac{P_{tx}(R)}{R^2} E[{x^2}] \nonumber
\end{eqnarray}
The nodes are randomly distributed over a sphere of radius R as shown in figure \ref{fig41}. The probability density of node is
\begin{eqnarray}
f_x(x)&=&\frac{3}{4\pi R^3} \nonumber \\
Hence &&\nonumber \\
E[{x^2}]&=& \int_{V} {x^2} f_x(x) dV \nonumber \\
E[x^2]&=& \int_V {x^2} \frac{3}{4\pi R^3} dV \nonumber \\
E[x^2]&=& \frac{3}{4\pi R^3} \int_V {x^2} dV \nonumber \\
E[{x^2}]&=& \frac{3}{4\pi R^3} \int_0^{\pi} \! \int_0^{2\pi} \! \int_Q^R  {x^4} \sin \theta dx d\theta d\phi \nonumber \\
E[{x^2}]&=& \frac{3 R^2}{5} \nonumber
\end{eqnarray}
Therefore,
\begin{eqnarray}
E[P_{tx}(x)] &=& \frac{3 P_{tx}(R)}{5} \nonumber \\
E[P_{pc}] &=& \xi [P_{tx}(R) \times T_c + P_{rx}\times[(C-1)T_c + \lambda T_d + \lambda T_a]] \nonumber \\
&&+ \xi (\lambda T_d + \lambda T_a) P_{tx}(R) \frac{3}{5} \nonumber
\end{eqnarray}
Now we calculate the amount of energy saved by substracting $E_{pc}$ from $E_{wpc}$.
\begin{eqnarray}
 P_{saved} &=& P_{wpc}-p_{pc} \nonumber \\
&=& \frac{2}{5} \xi \lambda P_{tx}(R)(T_d+T_a) + (D-2\lambda)\xi P_{rx}(T_d+T_a)
\end{eqnarray}

\begin{figure}[!t]
\centering
\includegraphics[scale=0.27]{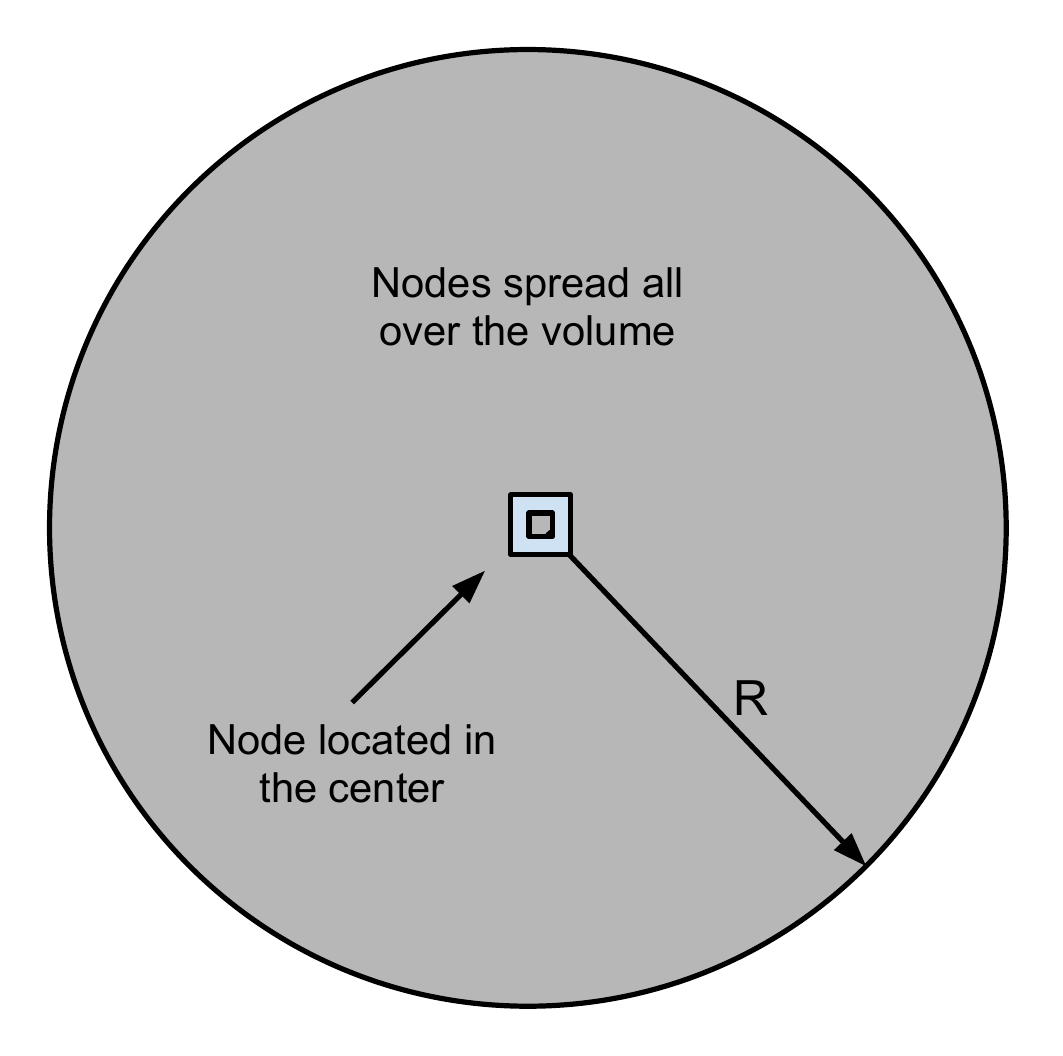} 
\caption{Nodes distribution over the volume of sphere of radius Q and R}
\label{fig41}
\end{figure}

From equation 4.11 we can see the amount of power saved. We use this equation to plot the graph of power saving vs number of nodes as shown in graph in figure \ref{graph5}. The various parameters taken is shown in table \ref{tab43}. We can see from the graph that the power saved increases to a maxima of about 0.46 Watts and then falls down gradually. Note that the graph is plotted for the optimal case. With the increasing number of users the CR nodes have to receive more control packets, hence the power requirements increase.
\begin{table}
\begin{center}
\caption{Values of various parameters for simulation in section \ref{enr}}
    \begin{tabular}{ | l | l |}
    \hline
   N & Varied from 4 to 20\\ \hline  
   $P_{tx}(R)$ & 1500 mW [19]\\ \hline
   $P_{rx}$ & 800 mW [19]\\ \hline 
  	$T_c$ & 1 msec\\ \hline 
  	$T_d$ & 2 msec\\ \hline
  	$T_a$ & 0.5 msec\\ \hline
  	$T_q$ & 20 msec\\ \hline
  	$T_s$ & 80 msec \\ \hline
    \end{tabular}
\end{center}
\label{tab43}
\end{table}

\subsection{Performance Evaluation in presence of a Malicious User}
\label{mal}
Malicious user in this context refers to a node which desires to always avail the data slot without caring for other users. A node can become selfish by manipulating its priority index in such a way that the packet scheduling algorithm (PSA) always grant him the data slot. In this case the other nodes which honestly want to transmit their data may not get a chance to transmit. To eliminate this problem we have used the past PSA output (PPSA) component to indicate whether the data slot was allocated to this CR node in the last superframe. Hence every node keeps track of this index (PPSA). This PPSA is added to the priority index (PI) to get net priority index (NPI).\\
\indent Suppose in a scenario a malicious node $A$ always requests for slot by declaring its PI as 21. There is a honest node $B$ which claims its PI to be 18. Let us assume that malicious node $A$ have large chunk of data which would require all the data slots for its transmission. During the first superframe as there was no data slot allocated to any node NPI = PI. Hence during the first superframe, node $A$ will be allocated the data slot while the node $B$ will not be allocated any slot. In the second superframe the PPSA would play a role. PPSA for node $A$ now is 5 (slot allocated in the last superframe) while that of node $B$ is 0 (slot not allocated in the last superframe). The net priority NPI=PI+PPSA for node $A$ is 21 while that of node $B$ is 23. Hence node $B$ will be allocated the data this time. This method reduces the chance of hogging the bandwidth by a particular malicious user and provide fair chance to other honest CR users to transmit their data.\\
\indent We provide the simulation result for the same. The values of various parameters are shown in table \ref{tab44}. There are 2 flows in the network, one from node 1 to node 3 and node 2 to node 4.  The simulation result shows that the throughput of malicious node $A$ is 306.36 Kbps and that of node $B$ is 136.62 kbps. the results show that the other nodes which try to send their packets honestly also get a fair chance to access the channel.
\begin{table*}
\begin{center}
\caption{Values of various parameters for simulation in section \ref{mal}}
    \begin{tabular}{ | l | l | l | l |}
    \hline
   Total Nodes & 6 & Packet size & 2000 bytes\\ \hline
   Data Slots & 15 & Packet generation Interval & 1 msec\\ \hline   
  	$T_c$ & 1.5 msec & Queue length & 100\\ \hline 
  	$T_d$ & 2 msec & Routing & AODV\\ \hline
  	$T_a$ & 0.5 msec & Simulation time & 10 sec\\ \hline
  	$T_q$ & 20 msec & Source & CBR\\ \hline 
Flow 1 & Node 1 to 3 & PI of node 1 & 21 \\ \hline
Flow 2 & Node 2 to 4 & PI of node 2 & 18\\ \hline
Throughput Node 1 & 306.36 kbps & Throughput Node 3 & 136.62 kbps \\ \hline
\end{tabular}
\end{center}
\label{tab44}
\end{table*}

\subsection{Fairness among CR nodes}
\label{fair}
In our MAC protocol the fairness is maintained in the sense that all the nodes get equal opportunity for a data slot if they have their priorities same. Even if the data priorities are different the priority index sent by a node also depends on the waiting time of the packet in queue. Hence at some time a non-real time data packet would get a high PI and hence could get a data slot among the real time data traffic.\\
\indent In table \ref{tab48} we give the details of simulation to investigate the fairness among the nodes by allocating a random PI to all the nodes. We have generated four flows among the 8 CR users. We compute the ratio of throughput of flow 1 Vs throughput of flow 2 Vs throughput of flow 3 Vs flow 4. If it comes out close to 1 then we can say that the protocol is fair hence the more it deviates from 1 the more protocol is unfair.\\
\indent According to the results the ratios are 1.012, 0.9477, 1.0281 and 1.012. Hence we can see that the ratios are close to 1 for all the nodes and hence proves that the protocol provides equal opportunity to all the nodes. 
\begin{table*}
\begin{center}
    \begin{tabular}{ | l | l | l | l |}
    \hline
   Total Nodes & 8 & Packet size & 2000 bytes\\ \hline
   Data Slots & 13 & Packet generation Interval & 1 msec\\ \hline   
  	$T_c$ & 1.5 msec & Queue length & 100\\ \hline 
  	$T_d$ & 2 msec & Routing & AODV\\ \hline
  	$T_a$ & 0.5 msec & Simulation time & 20 sec\\ \hline
  	$T_q$ & 20 msec & Source & CBR\\ \hline 
Flow 1 & Node 1 to 2 & Flow 3 & Node 5 to 6\\ \hline
Flow 2 & Node 3 to 4 & Flow 4 & Node 7 to 8\\ \hline
Throughput flow 1 & 89.79 kbps & Throughput flow 3 & 91.21 kbps \\ \hline
Throughput flow 2 & 84.08 kbps & Throughput flow 4 & 89.79 kbps \\ \hline
\end{tabular}
\end{center}
\caption{Values of various parameters for simulation in section \ref{fair}}
\label{tab48}
\end{table*}

\subsection{QoS performance of the protocol}
\label{qos}
In this section we would show the QoS performance of the protocol. As the data slots are assigned according to the priority index, hence the real time traffic with high PI is routed first as compared to low priority traffic. We simulate the network with the details as given in table \ref{table46}. The throughput is shown for each flow. As seen from the simulation result, that the high priority flow has high throughput and subsequently the low priority flow has lower throughput.
\begin{table*}
\begin{center}
    \begin{tabular}{ | l | l | l | l |}
    \hline
   Total Nodes & 8 & Packet size & 2000 bytes\\ \hline
   Data Slots & 17 & Packet generation Interval & 1 msec\\ \hline   
  	$T_c$ & 1.5 msec & Queue length & 100\\ \hline 
  	$T_d$ & 2 msec & Routing & AODV\\ \hline
  	$T_a$ & 0.5 msec & Simulation time & 20 sec\\ \hline
  	$T_q$ & 20 msec & Source & CBR\\ \hline 
Flow 1 & PI 21 & Flow 3 & PI 9\\ \hline
Flow 2 & PI 15 & Flow 4 & Pi 3\\ \hline
Throughput flow 1 & 130.03 kbps & Throughput flow 3 & 71.76 kbps \\ \hline
Throughput flow 2 & 81.71 kbps & Throughput flow 4 & 59.68 kbps \\ \hline
\end{tabular}
\end{center}
\caption{Values of various parameters for simulation in section \ref{qos}}
\label{tab46}
\end{table*}

\subsection{Improvements in protocol}
In this section we will propose some modifications in the protocol to improve its throughput and overall its efficiency.\\
\indent Every node in the network has the information about the distance of other nodes in the network. This is obtained by operating on the sent and received power to know the SNR. During the data transmission phase the nodes can use this SNR calculated to suitably modify the modulation scheme like QPSK, 16-QAM, 64-QAM for transmission over their allocated data slots. This increases the overall throughput of the system and achieves channel utilisation as the data transmitted per unit time slot increases. Also the SNR is also maintained which enables the system to combat against any channel deterioration. This mechanism enables the system to reconfigure itself according to the channel parameters.\\
\indent There could be scenarios in which PU activity is very frequent but the duration of PU activity is very small. In this case the lenght of superframe has to be made shorter so that the channel is sensed frequently. We propose to use multiple quiet periods in between the superframe. Subsequently the length of the superframe comprising of control sequence, data sequence and multiple quiet periods could be increased. The duration of superframe, as we stated earlier, is based on the pattern of PU activity. This is bounded by the quiet periods. Hence, when we introduce the quiet periods in between the length of superframe can now be made significantly larger. This technique enables us to cause less disturbance in PU operation as well as also allow the CR users to make full use of the available spectrum. This also reduces the overhead in control packet transmission as the control packets are transmitted once per superframe, where superframe length has increased significantly.\\
\indent The usability of the spectrum can further be increased by using directional antennas. During the control sequence phase where the control packets are broadcast over to all the nodes, we can use directional antenna to know the direction of transmission. At the end of this control sequence, all the nodes can have the knowledge of the topology with the distance and direction of other nodes. Hence during the data transmission phase the transmissions could be done in one direction i.e. direction towards the receiver. Hence this could enable the reuse of same data slot by other users geographically. This could increase the throughput tremendously. 

%%%%%%%%%%%%%%%%%%%%%%%%%%%%%%%%%%%%%%%%%%%%%%%%%%%
%%%%%%%%%%%%%%%\input{paperch5}%%%%%%%%%%%%%%%%%%%%%%%%%%
%%%%%%%%%%%%%%%%%%%%%%%%%%%%%%%%%%%%%%%%%%%%%%%%%%%
\section{Conclusion and Future Works}
In this article, we have tried to solve the link layer problem in the Cognitive
radio networks. We have proposed an adaptive TDMA based MAC protocol for
Cognitive radio Networks. We have discussed the various recent protocols
proposed in the literature and stated the limitations of these. This has
motivated us to develop a novel protocol which can surpass these limitations.
The adaptive TDMA MAC protocol divides time into time slots for communication of
control and data traffic. This does not require a dedicated control channel for
its operation, hence removes problems like channel saturation and availability
of such band. The significant contribution of the protocol apart from being
energy efficient is that it ensures fairness among nodes and maintain QoS.\\
\indent We have simulated the MAC protocol over ns-2 and compared it with the common control channel based MAC protocol. We have seen the effect on throughput with the variation in different parameters like quiet period duration, super frame duration and primary user activity. We have also talked about the resilience of our protocol in case of presence of a malicious user. We have also derived the expressions for power evaluation of our protocol and theoretical maximum throughput. Main results of our protocol are as follows:
\begin{itemize}
\item Protocol maintained QoS by allowing higher throughput for high priority data traffic and low throughput for subsequently low priority data traffic.
\item Performance in presence of malicious user was fair with a good amount of throughput ensured for low priority nodes.
\item Energy efficiency of the protocol is seen to be significant with a high amount of power being saved.
\item Our protocol ensured maximum use of the available spectrum with no part of the spectrum left without any transmission.
\end{itemize}
Along with the analysis and simulation of our MAC protocol we also proposed some improvements to reduced the overload in transmitting control packets and increase efficiency of our system. we also proposed schemes for high data transfer over the data slots.

\subsubsection{Commercial Possibilities}
Our protocol could be used efficiently in Cognitive Radio adhoc networks and Cognitive Radio Sensor networks. The protocol provides an optimality between throughput and energy efficiency. This protocol could be also be used in Cognitive Radio systems for home surveillance applications.

\subsection{Future Works}
\begin{itemize}
\item A routing protocol has to be developed in accordance with this protocol, so that the system becomes more efficient.
\item Study and further implementation of the proposed improvements in section 4.x are to be carried out.
\item Performance of the protocol has to be compared with various other protocols in the literature.
\item Evaluation and modification of the protocol for various primary network scenarios like TV White spaces, Radio Spectrum etc. need to be done. 
\end{itemize}

\end{document}